%% file: main.tex
\documentclass[sigconf]{acmart}

\usepackage{algorithm}
\usepackage{algorithmic}

\usepackage{cleveref}
\usepackage{stmaryrd}   
\usepackage{multirow}    
\usepackage{makecell}    
\usepackage{enumitem}    
\usepackage{pifont}      
\definecolor{step2color}{RGB}{192, 0, 0}      
\definecolor{step3color}{RGB}{255, 128, 0}    

\usepackage{xspace}

\newcommand{\method}{ROSETTA\xspace}

\copyrightyear{2026} 
\acmYear{2026} 
\setcopyright{cc} 
\setcctype{by} 
\acmConference[CCS '26]{Proceedings of the 2026 ACM SIGSAC Conference
  on Computer and Communications Security}{November 15--19, 2026}{The Hague, Netherlands} 
\acmBooktitle{Proceedings of the 2026 ACM SIGSAC Conference on Computer and
  Communications Security (CCS '26), November 15--19, 2026, The Hague, Netherlands}
\acmDOI{10.1145/3830454.3846549} 
\acmISBN{979-8-4007-2871-6/2026/11} 
\begin{document}

\title{ROSETTA: Efficient and Accurate Privacy-Preserving LLM Decoding via Hybrid CKKS/TFHE Evaluation}

\author{Jiangrui Yu}
\affiliation{%
  \institution{Peking University}
  \city{Beijing}
  \country{China}
}
\email{jiangrui.yu@stu.pku.edu.cn}

\author{Baosheng Zhang}
\authornote{This work was completed while Baosheng Zhang was an intern at Peking University.}
\affiliation{%
  \institution{Xi'an Jiaotong University}
  \city{Xi'an}
  \country{China}
}
\email{2831532315@stu.xjtu.edu.cn}

\author{Liang Kong}
\affiliation{%
  \institution{Ant Group}
  \city{Beijing}
  \country{China}
}
\email{kongliang.kong@antgroup.com}

\author{Lin Ding}
\affiliation{%
  \institution{Peking University}
  \city{Beijing}
  \country{China}
}
\email{dinglin@pku.edu.cn}

\author{Yi Chen}
\affiliation{%
  \institution{Peking University}
  \city{Beijing}
  \country{China}
}
\email{yichen25@stu.pku.edu.cn}

\author{Ye Yu}
\affiliation{%
  \institution{Peking University}
  \city{Beijing}
  \country{China}
}
\email{yy3628@columbia.edu}

\author{Mingzhe Zhang}
\affiliation{%
  \institution{Ant Group}
  \city{Beijing}
  \country{China}
}
\email{smartzmz@gmail.com}

\author{Meng Li}
\authornote{Corresponding author.}
\affiliation{%
  \institution{Peking University}
  \city{Beijing}
  \country{China}
}
\email{meng.li@pku.edu.cn}

\renewcommand{\shortauthors}{Jiangrui Yu et al.}

\input{docs/1-abstract}



\maketitle

\input{docs/2-introduction}

\input{docs/3-background}

\input{docs/4-design}

\input{docs/5-framework}

\input{docs/5-evaluation}

\input{docs/6-conclusion}

\begin{acks}
This work was supported in part by the
\grantsponsor{NSFC}{National Natural Science Foundation of China (NSFC)}
{https://doi.org/10.13039/501100001809}
under Grants \grantnum{NSFC}{92464104}, \grantnum{NSFC}{62495102}, and
\grantnum{NSFC}{62341407}; in part by the
\grantsponsor{NKRDP}{National Key Research and Development Program}{}
under Grant \grantnum{NKRDP}{2024YFB4505004}; in part by Ant Group through the
\grantsponsor{CCF-Ant}{CCF-Ant Research Fund}{}
under Grant \grantnum{CCF-Ant}{CCF2412628150};
in part by the
\grantsponsor{BAIC-FBPC}{Beijing Advanced Innovation Center for Future
Blockchain and Privacy Computing}{}
under Grant \grantnum{BAIC-FBPC}{GJJ-25-014}; and in part by the
\grantsponsor{111 Project}{111 Project}{}
under Grant \grantnum{111 Project}{B18001}.
\end{acks}

\bibliographystyle{ACM-Reference-Format}
\balance
\bibliography{reference/ref}

\appendix

\section{Open Science}
\label{appendix:open-science}
To promote availability, all pre-trained models used in this work
(GPT-2, TinyLlama-1.1B, LLaMA-3-8B, Qwen2-7B, Mistral-7B, and
DeepSeek-R1-Distill-LLaMA-8B) and the WikiText-2 evaluation corpus
are publicly accessible through their original sources, as
referenced in the paper. Our source code and materials for
replicating the experiments are publicly available at
\url{https://anonymous.4open.science/r/ROSETTA_artifact-B77F/}.
We welcome feedback and contributions to
improve the implementation and further the research in
privacy-preserving inference.

\section{Ethical Considerations}
\label{appendix:ethics}
This work focuses on improving the efficiency of privacy-preserving
inference frameworks for generative large language models. Our
contributions are intended to advance private computation techniques
without creating new risks to data security or user privacy. By
enhancing the practicality and scalability of privacy-preserving
machine learning, we hope to encourage the responsible and broader
uptake of privacy-centric technologies.

We strictly comply with the ACM Code of Ethics. In particular, our
work is guided by the principles of respecting user privacy
(Principle~1.6) and avoiding harm (Principle~1.2). All models and
datasets employed in our methodology, including GPT-2, TinyLlama-1.1B,
LLaMA-3-8B, Qwen2-7B, Mistral-7B, DeepSeek-R1-Distill-LLaMA-8B,
WikiText, LAMBADA, GSM8K, and ShareGPT, are publicly accessible. Our
experiments do not intentionally collect or process proprietary,
sensitive, or personally identifiable information. \method{} runs
entirely under FHE on ciphertext-aligned
activations and weights, and does not introduce additional privacy
risk beyond that of the underlying CKKS and TFHE primitives or the
threat model assumed by prior FHE-LLM systems. We have carefully
considered the ethical implications of our segmented LUTs and
scheme-aware operator selector, and confirm that no privacy
violations arise from our experimentation or proposed methods.

In summary, because this research does not involve human subjects and
does not intentionally collect or process proprietary, sensitive, or
personally identifiable information, we identify no need for formal
human-subject review.
By advancing the practicality of privacy-preserving machine learning,
we believe this work helps promote the responsible adoption of
privacy-enhancing technologies and reinforces data minimization and
confidentiality.

\section{Generative AI Disclosure}
\label{appendix:genai}
Generative AI tools such as GPT, Codex, and Gemini were employed to support code implementation, benchmark configuration, \LaTeX{} typesetting, and the editing and polishing of portions of the manuscript. All AI-assisted content, including prose, code, experimental results, and citations, was thoroughly inspected, validated, and revised by the authors, who assume full responsibility for the accuracy, originality, and integrity of the work presented in this paper.

\end{document}

%% file: docs/1-abstract.tex
\begin{abstract}
Generative large language models (LLMs) have achieved state-of-the-art performance on many real-world tasks such as code generation and question answering. These models predominantly rely on an autoregressive decoding strategy that generates output tokens sequentially. However, their pervasive deployment raises serious privacy concerns, motivating private inference frameworks based on fully homomorphic encryption (FHE). A major limitation of existing FHE frameworks is their inefficiency in evaluating nonlinear operations, which incur substantial overhead and dominate the decode stage.

In this paper, we propose ROSETTA, a hybrid CKKS/TFHE framework that overcomes this limitation. We first observe that nonlinear operations in the decode stage exhibit heterogeneous workload patterns, which can be handled effectively via a hybrid approach. We then realize this with two key contributions: 1) an adaptive segmented lookup-table protocol based on TFHE that enables efficient and accurate evaluation of nonlinear operations; and 2) a scheme-aware operator-selection framework that automatically assigns each nonlinear operator to CKKS or TFHE to minimize end-to-end decoding latency. We demonstrate that ROSETTA achieves up to 4.8$\times$ Softmax speedup and 1.5--2.1$\times$ end-to-end speedup over the SOTA framework CacheMir. 
\end{abstract}

%% file: docs/2-introduction.tex
\section{Introduction}
\label{sec:introduction}

Generative large language models (LLMs), such as GPT~\cite{radford2019gpt} and LLaMA~\cite{touvron2023llama2openfoundation}, have demonstrated impressive performance across a broad range of applications, such as clinical diagnostics~\cite{Shamshad2022TransformersIM}, financial analysis~\cite{ding2026largelanguagemodelagent}, document summarization~\cite{koh2022longdocsumm}, and intelligent voice assistants~\cite{Cheng2022PersonalVA}. These models typically execute on cloud platforms and process user prompts that may contain highly sensitive information. Consequently, privacy has become a fundamental concern for the deployment of LLM inference~\cite{hao2022iron,pang2023bolt,lu2023bumblebee,blb2025xu}.

\begin{figure}[t]
  \centering
  \includegraphics[width=\columnwidth]{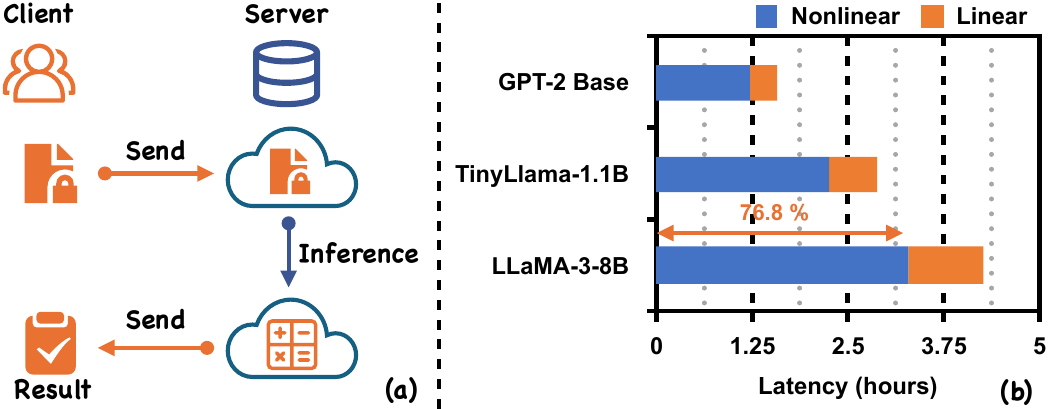}
  \Description{Two-panel overview. The left panel shows a client sending encrypted input to a server for private inference and receiving an encrypted result. The right panel compares decoding latency across three language models and shows that nonlinear operations account for most of the runtime.}
  \caption{(a) Illustration of FHE-based private inference. (b) Latency breakdown for GPT-2 Base, TinyLlama-1.1B, and LLaMA-3-8B based on CacheMir~\cite{yu2026cachemir}.}
  \label{fig:intro-profile}
\end{figure}

\begin{figure*}[t]
    \centering
    \includegraphics[width=\textwidth]{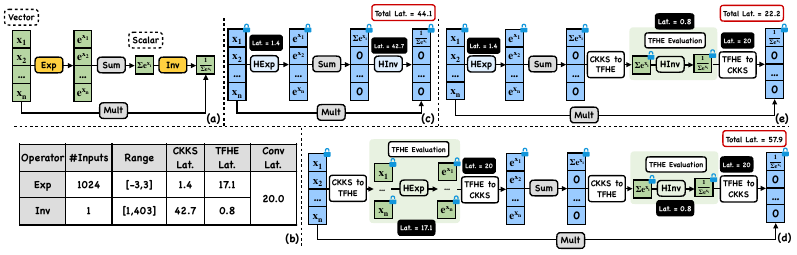}
    \Description{Five-panel comparison of softmax evaluation. The first two panels show the softmax pipeline and measured module costs. The remaining panels compare all-CKKS, all-TFHE, and hybrid CKKS/TFHE assignments, with the hybrid assigning exponential evaluation to CKKS and reciprocal evaluation to TFHE.}
    \caption{Softmax evaluation under different FHE schemes ($n{=}1024$). (a) Plaintext softmax pipeline. (b) Measured per-module latency under CKKS or TFHE. (c)--(e) Three schemes for evaluating the nonlinear operators: (c) all under CKKS, (d) all under TFHE, and (e) a hybrid assignment. ``Lat.''\ and ``Conv.''\ denote latency and scheme conversion, respectively.}
    \label{fig:softmax-decomposition}
\end{figure*}

To address this challenge, private inference frameworks have been proposed to protect both proprietary model weights and user inputs throughout inference. Specifically, the client learns nothing beyond the final output, while the server obtains no information about the input data. Existing systems have explored cryptographic frameworks based on fully homomorphic encryption (FHE)~\cite{chen2022thex,balcan2016cryptonets,ebel2025orion,kim2023convfhe,moon2025thor,park2024powerformer,zhang2025moai,zhang2024nexus,yu2026peft,deng2024trinity}, secure multi-party computation (MPC)~\cite{Akimoto2023PrivformerPT,dong2025puma,kanav2023sigma,ndss2025Kei,li2023mpcformer,zeng2023mpcvit,zeng2025mpcache,zeng2024eqo}, and hybrid FHE/MPC designs~\cite{hao2022iron,huang2022cheetah,juvekar_gazelle_2018,lu2023bumblebee,Mishra2020delphi,pang2023bolt,xu2024privcirnet,yu2024flexhe,zhang2025fenix,xu2023falcon,xu2024hequant,xu2024privquant,zhou2025cryptomoe}. Among these designs, FHE-based inference achieves higher communication efficiency and significantly reduces the computational burden on the client. As a result, FHE-centric approaches are more practical for end users: the client simply uploads encrypted inputs, the untrusted server carries out all subsequent computations directly over ciphertexts, and only the encrypted output is sent back, as shown in Figure~\ref{fig:intro-profile}(a).

Despite these benefits, building a fully end-to-end LLM over FHE remains substantially challenging. In particular, generative LLM inference generally involves two main stages: prefill and decode. The prefill stage processes the entire input prompt, typically hundreds to thousands of tokens, in parallel. This naturally fits SIMD-style schemes like CKKS~\cite{Cheon2018CKKS} and has been well optimized by pure-CKKS frameworks~\cite{moon2025thor,zhang2025moai,zhang2024nexus,park2024powerformer}. The decode stage, in contrast, processes only one token at a time, potentially yielding low data parallelism for some operators, which can leave many SIMD slots underutilized with pure-CKKS frameworks. Although CacheMir~\cite{yu2026cachemir} has optimized its linear components, nonlinear operators remain the dominant source of latency, accounting for more than \textbf{70\%} of the decode stage total runtime, as illustrated in Figure~\ref{fig:intro-profile}(b).

To meet such disparate demands, a natural idea is to leverage SISD-style schemes like TFHE~\cite{chillotti2020tfhe} for nonlinear operators. TFHE supports programmable bootstrapping (PBS), a lightweight and flexible primitive that evaluates arbitrary functions through a look-up table (LUT). Operating on a per-element basis, PBS does not rely on SIMD batching and remains efficient when nonlinear operators receive only a few elements. Hybrid CKKS/TFHE frameworks such as PEGASUS~\cite{lu2021pegasus} explore this direction by offloading \textbf{all} nonlinear layers to TFHE: they first convert CKKS ciphertexts to TFHE, evaluate all nonlinear operators via PBS, and finally convert the results back to CKKS for subsequent linear layers.

However, we observe that nonlinear operators in the decode stage are highly heterogeneous along two dimensions, \textbf{element count} and \textbf{input dynamic range}, and a one-size-fits-all approach is suboptimal. Take softmax as an example (Figure~\ref{fig:softmax-decomposition}(a) and (b)): it comprises two nonlinear operators, $\exp$ and $1/x$. The $\exp$ operator processes $n$ elements with a bounded input range, whereas the subsequent summation produces a single scalar with a much wider range fed into $1/x$. Evaluating both operators under CKKS, as in pure-CKKS frameworks~\cite{moon2025thor,zhang2024nexus,zhang2025moai,yu2026cachemir}, handles $\exp$ efficiently with a low-degree polynomial approximation under SIMD batching (Figure~\ref{fig:softmax-decomposition}(c)). However, to achieve enough precision over the wide dynamic range of $1/x$, iterative methods such as Goldschmidt require many rounds, each consuming multiplicative levels and triggering frequent bootstrapping, yielding a total latency of 44.1s.

Evaluating both operators under TFHE instead (Figure~\ref{fig:softmax-decomposition}(d)) faces two fundamental issues. \emph{First}, $\exp$ over $n$ elements is inefficient because TFHE's element-wise PBS cannot exploit batch parallelism, and the two scheme conversions add further overhead, incurring 57.9s in total. \emph{Second}, although PBS is efficient on $1/x$, whose input is a scalar, a single PBS encodes only a small LUT (e.g., $2{,}048$ entries), capping precision at $\sim$11 bits, which is insufficient for the wide dynamic range and can inflate the softmax perplexity by over $100\%$ relative to the plaintext baseline (Table~\ref{tab:ppl}), rendering the efficient computation meaningless. \emph{In summary, neither scheme alone is sufficient. CKKS achieves high precision but at high cost and is restricted to wide SIMD batching, whereas TFHE flexibly evaluates per-element workloads but cannot deliver high precision.}

Therefore, their complementarity suggests a per-operator hybrid. As shown in Figure~\ref{fig:softmax-decomposition}(e), selectively offloading only $1/x$ to TFHE while keeping $\exp$ under CKKS combines the strengths of both schemes and reduces the total latency to 22.2s. However, realizing this idea raises two challenges: \underline{\textbf{(i)}} the TFHE side still inherits PBS's precision limitation on wide-range operators such as $1/x$; and \underline{\textbf{(ii)}} the per-operator scheme assignment is itself non-trivial: offloading an operator to TFHE not only changes its own latency, but also introduces scheme conversion overhead, while the levels of the surrounding CKKS layers must be set accordingly. It is therefore insufficient to just compare each operator's own latency across the two schemes alone to get the optimal assignment. Existing frameworks ignore this and resort to coarse heuristics, such as offloading all nonlinear layers (PEGASUS) or only the first $k$ layers (LOHEN~\cite{lohen2024}), yielding suboptimal performance. Both challenges must be resolved before selective offloading can deliver efficiency and precision. Table~\ref{tab:comparison} summarizes the limitations of existing frameworks along these dimensions.

\subsection{Contribution}
To resolve the above challenges, we propose \method{}, a hybrid CKKS/TFHE framework for efficient and accurate privacy-preserving LLM decoding, with two techniques each targeting one of the above challenges.

\noindent \ding{182}~\textbf{Adaptive Segmented LUT Protocol.} To address challenge (i), we propose an Adaptive Segmented LUT Protocol that delivers high-precision nonlinear evaluation while remaining lightweight under TFHE. The protocol partitions the input domain into non-uniform segments whose boundaries adapt to the function's local behavior, with each segment carrying its own LUT and piecewise-linear interpolation, concentrating LUT entries where the function varies most rapidly. To evaluate the LUT homomorphically, we design a three-step protocol that (1) compares the encrypted input against every segment boundary and sums the homomorphic comparison results to derive the encrypted segment index, (2) multiplies the encrypted input with precomputed coefficient polynomials and uses a PBS to derive the interval index, and (3) similarly multiplies the input with packed slope/offset polynomials and uses PBS to produce the linear evaluation. The protocol relies only on RLWE plaintext--ciphertext multiplications and PBS lookups, avoiding ciphertext--ciphertext fixed-point arithmetic that TFHE does not natively support, and delivers $12$--$19$ bit precision over wide input ranges, exceeding the $\sim$11-bit ceiling of single-PBS approaches by $6$--$10$ bits and running $8$--$9\times$ faster than CKKS-based iterative methods (\S\ref{sec:design}).

\noindent \ding{183}~\textbf{Scheme-Aware Operator Selection.} To address challenge (ii), we develop a framework that jointly optimizes per-operator scheme assignment and CKKS multiplicative-level allocation, extending CacheMir's pure-CKKS shortest-path formulation~\cite{yu2026cachemir} to scheme-aware decisions. Our method is driven by two key observations. First, each nonlinear layer decomposes into a mix of arithmetic and non-arithmetic primitives, and only non-arithmetic primitives (e.g., $\exp$, $1/x$) benefit from TFHE evaluation. We therefore restrict scheme assignment to non-arithmetic primitives, dramatically shrinking the search space while leaving CKKS-friendly arithmetic operations untouched. Second, the cost of every CKKS--TFHE conversion can be expressed as an edge weight on the existing level-allocation DAG, so injecting a level-0 TFHE node alongside each non-arithmetic primitive turns the joint optimization into a single shortest-path problem on the augmented DAG. The shortest path simultaneously identifies which non-arithmetic primitives to offload and the optimal CKKS level for every layer, yielding a globally optimal schedule without separate scheme-selection and level-allocation passes (\S\ref{sec:framework}).

\begin{table}[t]
  \centering
  \caption{\method{} vs.\ prior FHE-based nonlinear evaluation frameworks. ``Poly.''\ denotes polynomial approximation; ``Seg.\ LUT'' denotes segmented LUT.}
  \label{tab:comparison}
  \footnotesize
  \setlength{\tabcolsep}{4pt}
  \begin{tabular}{@{}l|cccc@{}}
    \toprule
    \textbf{Feature} &
      \makecell{\textbf{Pure-CKKS}\\ \scriptsize\cite{zhang2024nexus,moon2025thor,yu2026cachemir}} &
      \makecell{\textbf{PEGASUS}\\ \scriptsize\cite{lu2021pegasus}} &
      \makecell{\textbf{LOHEN}\\ \scriptsize\cite{lohen2024}} &
      \makecell{\textbf{\method{}}\\ \scriptsize(Ours)} \\
    \midrule
    Scheme               & CKKS     & Hybrid     & Hybrid    & Hybrid                     \\
    \midrule
    Evaluator            & Poly.    & Single LUT & Single LUT& Seg.\ LUT + Poly.          \\
    \midrule
    TFHE precision       & ---      & Low        & Low       & High                       \\
    \midrule
    Workload size        & Large    & Small      & Small     & Both                       \\
    \midrule
    Selection strategy   & All CKKS & All TFHE   & First-$k$ & Per-op$^{*}$               \\
    \bottomrule
  \end{tabular}
  \\
  {\scriptsize $^{*}$ Jointly optimises per-operator scheme assignment and CKKS-level allocation.}
\end{table}

\noindent Overall, we make the following contributions:
\begin{itemize}[leftmargin=*]
  \item We design an adaptive segmented LUT protocol that achieves high-precision nonlinear evaluation over wide input ranges (\S\ref{sec:design}).
  \item We formulate the joint optimization of TFHE operator selection and CKKS level assignment as a shortest-path problem on a DAG, enabling the proper FHE scheme selection of operators (\S\ref{sec:framework}).
  \item We evaluate \method{} against the SOTA pure-CKKS method CacheMir~\cite{yu2026cachemir}, and hybrid CKKS/TFHE method PEGASUS~\cite{lu2021pegasus}. Compared to CacheMir, \method{} achieves up to $\mathbf{4.8\times}$ Softmax speedup and $\mathbf{1.5}$--$\mathbf{2.1\times}$ end-to-end speedup. Compared to PEGASUS, \method{} achieves $\mathbf{3}$--$\mathbf{4\times}$ end-to-end speedup while reducing average PPL degradation from $+116\%$ to only $\mathbf{+1.4\%}$.
  
\end{itemize}

%% file: docs/3-background.tex
\section{Background}
\label{sec:background}

\subsection{Notation}
\label{sec:notation}

We use bold lower-case letters to represent vectors (e.g., $\mathbf{m}$),
with $\mathbf{m}[i]$ denoting the $i$-th element, and hatted lower-case
letters for polynomials (e.g., $\hat{m}(X)$).
We write $R_{N,Q} = \mathbb{Z}_Q[X]/(X^N+1)$ for the $2N$-th cyclotomic
ring modulo $Q$, and $\mathbb{Z}_q^{n_{\mathrm{LWE}}+1}$ for an $(n_{\mathrm{LWE}}+1)$-dimensional
vector modulo $q$.

\subsection{FHE Schemes}
\label{sec:bg-fhe}

In this work, we primarily use two types of FHE schemes: CKKS~\cite{cheon2017homomorphic}
and TFHE~\cite{chillotti2020tfhe}, which can both be instantiated over standard
lattice hardness problems.
In what follows, we sketch the main operators for each scheme.

\subsubsection{The CKKS Scheme}
The CKKS scheme is constructed over the Ring Learning with Errors ($\mathsf{RLWE}$) problem.
It natively encodes and encrypts multiple data elements (referred to as a \emph{packed} ciphertext) into a single $\mathsf{RLWE}$ ciphertext polynomial.
Here, we summarize the key operators over CKKS used in this paper.

\begin{itemize}[leftmargin=*]
\item \textbf{Encryption and Decryption:} Given a vector of plaintext messages $\mathbf{m} \in \mathbb{C}^{N/2}$, we have
\begin{equation}
    \llbracket \mathbf{m} \rrbracket_\mathsf{CKKS} \leftarrow \mathsf{CKKS\text{-}Enc}(\mathbf{m}),
\end{equation}
where $\llbracket \mathbf{m} \rrbracket_\mathsf{CKKS} \in R_{N,Q}^2$~\cite{cheon2017homomorphic}, for $N$ the ring dimension and $Q$ the ciphertext modulus, and $\mathbf{m} \approx \mathsf{CKKS\text{-}Dec}(\llbracket \mathbf{m} \rrbracket_\mathsf{CKKS})$. We omit how $\mathbf{m}$ is encoded into polynomials, as these details are orthogonal to our protocol.

\item \textbf{Addition and Multiplication:} $\mathsf{CKKS}$ ciphertexts
are compatible with element-wise ciphertext addition and multiplication, e.g.,
\begin{equation}
    \llbracket \mathbf{m}_0 + \mathbf{m}_1 \rrbracket_\mathsf{CKKS} = \mathsf{CKKS\text{-}Add}(\llbracket \mathbf{m}_0 \rrbracket_\mathsf{CKKS}, \llbracket \mathbf{m}_1 \rrbracket_\mathsf{CKKS}).
\end{equation}

\end{itemize}

\subsubsection{TFHE Scheme}
\begin{figure}[!t]
    \centering
    \includegraphics[width=\linewidth]{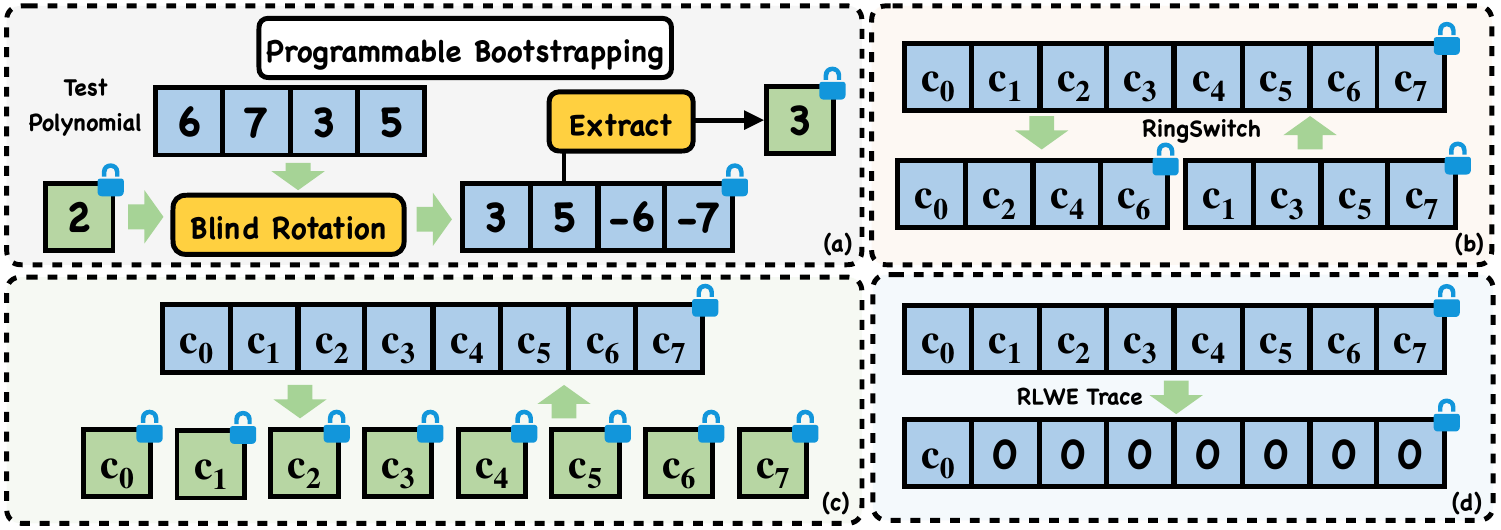}
    \Description{A block diagram depicting four fundamental ciphertext operations: (a) Programmable Bootstrapping, (b) Ring Switch, (c) LWE Extract and RLWE Repack, and (d) RLWE Trace Computation.}
    \caption{Operations used in our protocol: (a)~PBS, (b)~ring switch, (c)~LWE extract / RLWE repack, (d)~RLWE trace.}
    \label{fig:rlwe_primitives}
\end{figure}

The TFHE scheme is primarily constructed over the standard Learning with Errors ($\mathsf{LWE}$) problem.
It natively encrypts a single scalar data element into an $\mathsf{LWE}$ ciphertext.
It consists of the following operators:

\begin{itemize}[leftmargin=*]
\item \textbf{Encryption and Decryption:} Given a plaintext message $m \in \mathbb{Z}_q$, we have
\begin{equation}
    \llbracket m \rrbracket_\mathsf{TFHE} \leftarrow \mathsf{TFHE\text{-}Enc}(m),
\end{equation}
where $\llbracket m \rrbracket_\mathsf{TFHE} \in \mathbb{Z}_q^{n_{\mathrm{LWE}}+1}$~\cite{chillotti2020tfhe} is an LWE ciphertext. Here, $n_{\mathrm{LWE}}$ denotes the LWE dimension, and $q$ denotes the ciphertext modulus. Decryption recovers $m = \mathsf{TFHE\text{-}Dec}(\llbracket m \rrbracket_\mathsf{TFHE})$.

\item \textbf{Addition, Subtraction, and Constant Multiplication:}
$\mathsf{TFHE}$ ciphertexts are compatible with standard ciphertext addition,
subtraction, and constant multiplication. For brevity, we take only the
addition operator as an example, where it holds that
\begin{equation}
    \llbracket m_0 + m_1 \rrbracket_\mathsf{TFHE} = \mathsf{TFHE\text{-}Add}(\llbracket m_0 \rrbracket_\mathsf{TFHE}, \llbracket m_1 \rrbracket_\mathsf{TFHE}).
\end{equation}

\item \textbf{Programmable Bootstrapping (PBS)~\cite{chillotti2020tfhe}:}
PBS can homomorphically evaluate a look-up table (LUT) $f: \mathbb{Z}_q \to \mathbb{Z}_q$
while simultaneously refreshing the ciphertext noise:
\begin{equation}
    \llbracket f(m) \rrbracket_\mathsf{TFHE} = \mathsf{TFHE\text{-}LUT}_f(\llbracket m \rrbracket_\mathsf{TFHE}).
\end{equation}
This relies on a core primitive known as \emph{blind rotation}. To evaluate the LUT $f$,
its entries are encoded directly into the coefficients of a \emph{test polynomial} $\hat{v}(X) \in R_{N_\mathrm{br},Q_\mathrm{br}}$.
Here, $N_\mathrm{br}$ denotes the ring degree (which corresponds exactly to the total number of LUT entries),
and $Q_\mathrm{br}$ represents the ciphertext modulus. The blind rotation mechanism utilizes
the encrypted $\mathsf{TFHE}$ message $m$ as a shift amount to cyclically left-rotate
the test polynomial. This operation yields an $\mathsf{RLWE}^{N_\mathrm{br}}_{Q_\mathrm{br}}$
encryption of the shifted polynomial (where $\mathsf{RLWE}$ denotes a
standard Ring-LWE ciphertext). By extracting the constant term of it,
we obtain a fresh $\mathsf{TFHE}$ ciphertext encrypting the target value $f(m)$.

This blind rotation mechanism is illustrated in Figure~\ref{fig:rlwe_primitives}(a).
As shown, the LUT entries $\{f(0), f(1), \dots\}$ are laid out as the polynomial
coefficients (e.g., $6, 7, 3, 5$). The blind rotation cyclically shifts these coefficients
according to the encrypted index $m$. For instance, given an encrypted value $m=2$,
the polynomial is left-rotated by $2$ positions, moving the target entry $f(2) = 3$
directly into the constant-term position for subsequent extraction.
\end{itemize}

\subsubsection{Scheme Conversion and Other Primitives}
To change the underlying FHE scheme during the process of homomorphic evaluation, scheme conversion
operators seamlessly bridge packed RLWE ciphertexts and scalar LWE ciphertexts
~\cite{lu2021pegasus, boura2020chimera}. Below, we introduce the conversion operations utlized in our framework.

\begin{itemize}[leftmargin=*]
\item \textbf{Ring Switch~\cite{gentry2013field, bae2023hermes}:} As shown in Figure~\ref{fig:rlwe_primitives}(b), a ring switch is an operation that performs transitions between rings with different degrees $N$ and $N_{\mathrm{br}}$ (large/small rings).
The conversion can be processed in both directions by splitting one large ciphertext into multiple small ones, or the other way around. In our protocol, we specifically adopt the ring switching methodology from ~\cite{gentry2013field}.

\item \textbf{LWE Extract and RLWE Repack:~\cite{chillotti2020tfhe,chen2021ringconv}}
As depicted in Figure~\ref{fig:rlwe_primitives}(c), \emph{LWE Extract}~\cite{chillotti2020tfhe} extracts each coefficient of a packed $\mathsf{RLWE}$ ciphertext into independent $\mathsf{LWE}$ scalar ciphertexts, and \emph{RLWE Repack}~\cite{chen2021ringconv} performs the inverse, packing scalar LWE ciphertexts back into a single $\mathsf{RLWE}$ ciphertext.

\item \textbf{RLWE Trace Computation~\cite{chen2021ringconv}:} As illustrated in Figure~\ref{fig:rlwe_primitives}(d), the trace operation isolates the constant term $c_0$ of an $\mathsf{RLWE}$ ciphertext encrypting $a(X) = \sum_{i=0}^{N-1} c_i X^i$. Through a logarithmic sequence of Galois automorphisms, homomorphic additions, and a multiplication by $N^{-1}$, all non-constant terms cancel, leaving a ciphertext that holds only $c_0$.
\end{itemize}

\subsection{Modern LLM Architecture}
\label{sec:bg-llm}

\begin{figure}[htb]
    \centering
    \includegraphics[width=\linewidth]{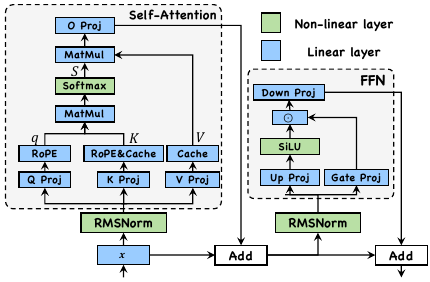}
    \Description{A block diagram illustrating the decode stage of a modern LLM architecture, showing the self-attention block (with Q/K/V projections, RoPE, KV-cache, and Softmax) and the FFN layer (with SwiGLU activation).}
    \caption{A modern Transformer block in LLaMA~\cite{touvron2023llama2openfoundation}.}
    \label{fig:llm_architecture}
\end{figure}

Modern LLM architecture is based on the Transformer~\cite{vaswani2017attention}, which consists of two components: the self-attention block and the feed-forward network (FFN). In this work, we focus on the decode stage of LLM and give a brief introduction below. We denote the model's hidden dimension as $d_\mathrm{model}$ (e.g., $4096$), the number of attention heads as $h$ (e.g., $32$), the per-head dimension as $d = d_\mathrm{model}/h$ (e.g., $128$), the FFN intermediate dimension as $d_\mathrm{ffn}$ (e.g., $11008$), and the current context length as $n$.

\paragraph{Self-Attention Block.}
Given a single-token input $\mathbf{x} \in \mathbb{R}^{1 \times d_\mathrm{model}}$, it is multiplied with three weight matrices to produce the query, key, and value vectors: $q = \mathbf{x} W_Q$, $k = \mathbf{x} W_K$, and $v = \mathbf{x} W_V$, all in $\mathbb{R}^{1 \times d}$. We refer to such operation, where the activation is multiplied by a weight matrix, as a \emph{projection}. The new key and value are appended to the KV-cache from previous tokens to form $K \in \mathbb{R}^{n \times d}$ and $V \in \mathbb{R}^{n \times d}$. The self-attention is then computed as:
\begin{equation}
    \mathbf{s} = \mathsf{Softmax}\!\left(\frac{q K^\top}{\sqrt{d}}\right) \in \mathbb{R}^{1 \times n}, \quad \mathbf{att} = \mathbf{s} \times V \in \mathbb{R}^{1 \times d}
\end{equation}
The attention output $\mathbf{att}$ from all $h$ heads is concatenated and passed through an output projection $W_O \in \mathbb{R}^{d_\mathrm{model} \times d_\mathrm{model}}$ to obtain the final output of the attention block.

\paragraph{FFN} The FFN layer is implemented using the SwiGLU activation. The input $\mathbf{x} \in \mathbb{R}^{1 \times d_\mathrm{model}}$ is first projected through both the up projection $W_\mathrm{up} \in \mathbb{R}^{d_\mathrm{model} \times d_\mathrm{ffn}}$ and the gate projection $W_\mathrm{gate} \in \mathbb{R}^{d_\mathrm{model} \times d_\mathrm{ffn}}$, followed by a SiLU activation, and finally passed through the down projection $W_\mathrm{down} \in \mathbb{R}^{d_\mathrm{ffn} \times d_\mathrm{model}}$ to produce the output:
\begin{equation}
    \mathsf{FFN}(\mathbf{x}) = \Big( \mathsf{SiLU}(\mathbf{x} W_\mathrm{up}) \odot (\mathbf{x} W_\mathrm{gate}) \Big) W_\mathrm{down}
\end{equation}

\paragraph{Max-free Softmax.} Let $\mathbf{z}$ denote the softmax input. Since the standard $\max$-subtraction is expensive in FHE, recent FHE frameworks~\cite{moon2025thor,zhang2024nexus,yu2026cachemir} adopt a \emph{normalize-and-square} formulation: starting from $y_i = e^{z_i/2^k}$ on a small range, the identity $\mathsf{Softmax}(2\mathbf{x})_i = y_i^2/\sum_j y_j^2$ is applied $k$ times to recover $\mathsf{Softmax}(\mathbf{z})$. We follow this formulation throughout.

\subsection{Threat Model}
\label{sec:bg-threat}

\method{} operates in a standard two-party private inference setting with a server and a client. The server hosts a proprietary LLM with confidential model weights, while the client holds sensitive input data~\cite{ju2024neujeans, mpcnn2022lee, hemet2021lou}. The protocol guarantees that the client obtains correct inference outputs while keeping both the server's model weights and the client's input private. Following prior work~\cite{ju2024neujeans, mpcnn2022lee, hemet2021lou}, we assume the model architecture is publicly known to both parties and operate under the honest-but-curious security model, wherein both parties adhere to the prescribed protocol but may attempt to learn additional information beyond what is permitted.

%% file: docs/4-design.tex
\section{Segmented LUT Evaluation}

\label{sec:design}
The observations from \S\ref{sec:introduction} drive us to design the \method{}~Segmented LUT protocol, which enables accurate and efficient nonlinear function evaluation through a lightweight LUT-based method. We first present the limitations of the single-LUT approach and introduce the Adaptive Segmented LUT that achieves high precision efficiently (\S\ref{sec:adaptive-segmented-lut}). We then realize this construction as a cryptographic lookup protocol (\S\ref{sec:online-execution-protocol}), which is seamlessly integrated into the LLM decoding flow via FHE scheme conversion (\S\ref{sec:pipeline-integration}).

\subsection{Adaptive Segmented LUT Construction}
\label{sec:adaptive-segmented-lut}

\begin{figure}[t]
    \centering
    \includegraphics[width=\columnwidth]{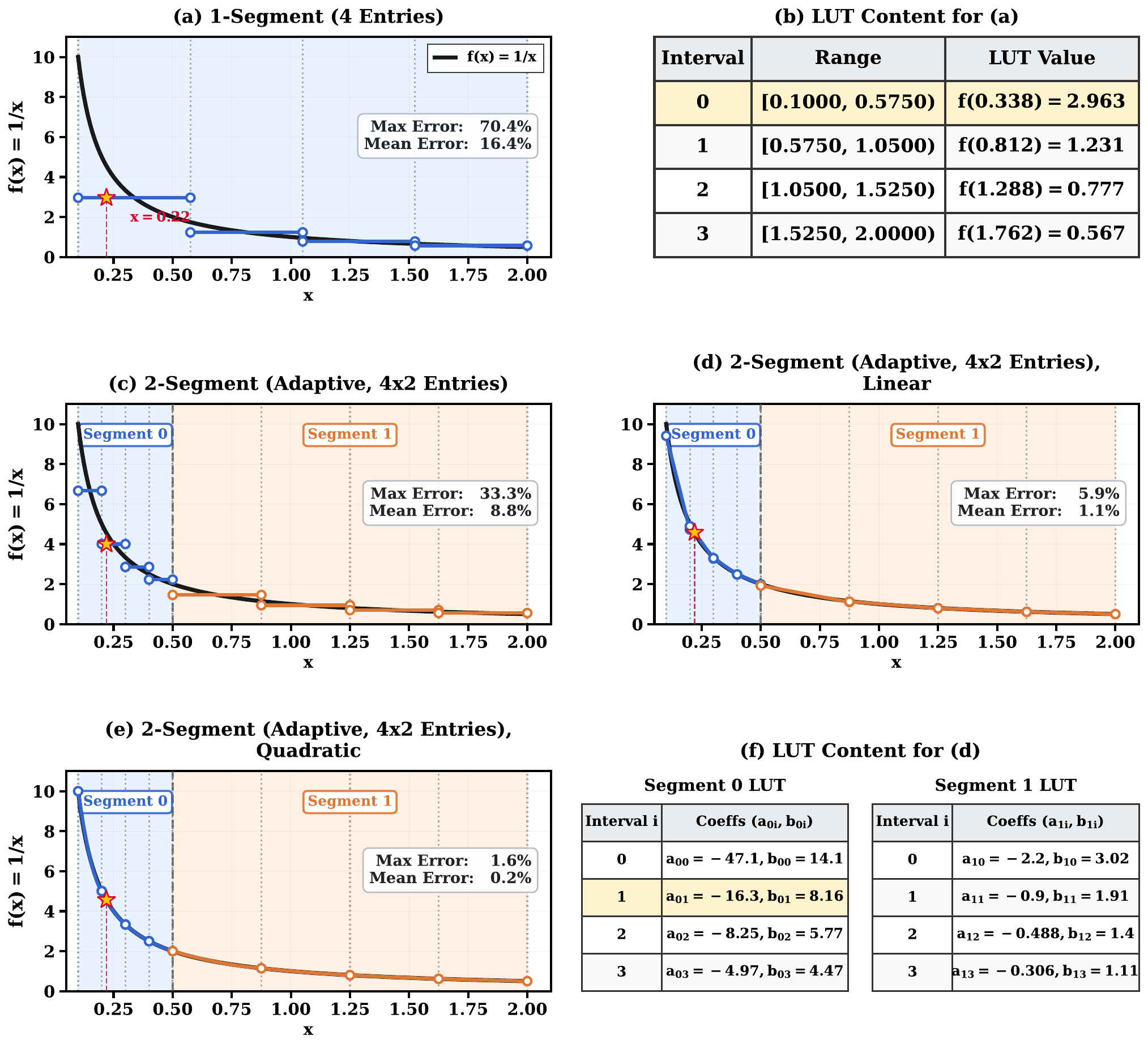}
    \Description{Six-panel illustration of approximating the reciprocal function with lookup tables. It contrasts a single uniform segment with two adaptive segments using constant, linear, and quadratic approximations, and lists the slope and offset coefficients for the linear construction.}
    \caption{Segmented LUT for $f(x){=}1/x$: (a)~single-segment baseline with (b)~its table; (c)--(e)~two-segment variants using constant/linear/quadratic approximations; (f)~slope/offset coefficients $(a_{si},b_{si})$ of the linear variant in (d).}
    \label{fig:segmented-lut}
\end{figure}

We start with the conventional single-LUT approach, which corresponds to the capability of a single TFHE PBS operation. Throughout this section, we use the evaluation of $f(x) = 1/x$ over the input domain $[\tau_{\mathrm{start}}, \tau_{\mathrm{end}}] = [0.1, 2.0]$ via a LUT with $N_{\mathrm{br}} = 4$ entries as a running example. As illustrated in Figures~\ref{fig:segmented-lut}(a) and (b), a standard LUT divides the domain into $N_{\mathrm{br}}$ equal intervals, approximating the function within each interval by its value at the midpoint. For instance, the first interval $[0.1, 0.575)$ has a midpoint of $0.338$; thus, $f(0.338) = 2.963$ is stored as the first LUT entry. 

To evaluate an input such as $x = 0.22$, we first determine which interval it falls into, denoted by index $I$. Because the intervals are equally sized, the index is proportional to the input's distance from $\tau_{\mathrm{start}}$. Thus, we directly compute $I$ via a linear mapping:
    \begin{equation*}
        I = \left\lfloor \frac{x - \tau_{\mathrm{start}}}{\tau_{\mathrm{end}} - \tau_{\mathrm{start}}} \cdot N_{\mathrm{br}} \right \rfloor = \left\lfloor \frac{0.22 - 0.1}{2.0 - 0.1} \cdot 4 \right \rfloor = 0
    \end{equation*}
Using this index ($I=0$), we fetch the pre-computed value from the table. As shown by the yellow row in Figure~\ref{fig:segmented-lut}(b), the returned result is the midpoint approximation $f(0.338) = 2.963$. Overall, this basic LUT method produces a ``staircase'' approximation of the original function $f(x)$. 

However, this single-LUT approach suffers from two main weaknesses. First, constrained by the limited capacity, a single table partitions the entire domain into only $N_{\mathrm{br}}$ equally spaced intervals, making each interval too wide to accurately capture rapid function variations. Second, within each interval, the function is approximated by a single constant (the midpoint value), which is accurate only near the midpoint; inputs far from it suffer from large errors. Together, these constraints significantly degrade approximation accuracy. As illustrated in Figure~\ref{fig:segmented-lut}(a), the baseline LUT evaluates $f(0.22)$ to only $2.963$, whereas the exact value is $4.545$. This yields a $34.8\%$ relative error at this evaluation point, and the maximum error reaches $70.4\%$ across the entire domain.

\subsubsection{Adaptive LUT Construction}
 To address the two aforementioned weaknesses, the table construction is optimized across two dimensions. First, to overcome the capacity limitation, an adaptive multi-segment LUT approach is adopted, as illustrated in Figure~\ref{fig:segmented-lut}(c). Rather than uniformly spanning the entire domain with a single table, the domain is adaptively partitioned into $N_{\mathrm{seg}} = 2$ segments, delimited by $N_{\mathrm{seg}}+1$ boundaries $\tau_0 = \tau_{\text{start}} < \tau_1 < \dots < \tau_{N_{\mathrm{seg}}} = \tau_{\text{end}}$, with a dedicated LUT assigned to each segment. In our example, the domain $[0.1, 2.0]$ is split into a narrow first segment $[\tau_0, \tau_1] = [0.1, 0.5]$ and a broader second segment $[\tau_1, \tau_2] = [0.5, 2.0]$. This non-uniform partition is because the function $f(x) = 1/x$ is steep near zero, necessitating a higher density of entries to maintain accuracy, whereas for larger inputs the curve flattens out and requires far fewer entries. By adapting the boundary spacing to the function's curve, the construction concentrates entries where the function varies most rapidly and reduces the maximum error to $33.3\%$.

Second, to mitigate the inaccuracies of a constant-value representation, each interval approximates $f(x)$ by a linear function $f(x) \approx a_{si} \cdot x + b_{si}$, where the slope $a_{si}$ and offset $b_{si}$ are precomputed and stored for interval $i$ of segment $s$ (Figure~\ref{fig:segmented-lut}(d, f)). For example, given $x = 0.22$ in segment $s=0$, interval $i=1$ (i.e., $[0.2, 0.3)$), the stored coefficients are $a_{01} = -16.327$ and $b_{01} = 8.163$, yielding $f(0.22) \approx -16.327 \times 0.22 + 8.163 = 4.571$, close to the true value $4.545$. Compared to the constant-value baseline, this linear model tracks the curve continuously within each interval, reducing the maximum error to $5.9\%$. While higher-order models such as quadratic approximation can further improve precision (Figure~\ref{fig:segmented-lut}(e)), we adopt the linear variant as it delivers sufficient accuracy at minimal cost.

\subsubsection{Online Lookup Evaluation.} After constructing the adaptive LUT offline, an online lookup algorithm is required to evaluate the table for a given query $x$. We first demonstrate a cleartext algorithm and extend it to the fully encrypted version in \S\ref{sec:online-execution-protocol}. The lookup proceeds by first locating which segment and interval the query $x$ falls in, denoted by the segment index $S$
and local interval index $I$, then retrieving the corresponding coefficients and performing a linear evaluation. Accordingly, the workflow is divided into three steps:
1) Segment Index Lookup, 2) Local Interval Index Computation, and 3) Coefficient Lookup and Linear Evaluation.

\textbf{Step 1: Segment Index Lookup.} Given an input $x$, the segment index $S \in \{0, \dots, N_{\mathrm{seg}}-1\}$ is derived by comparing $x$ against each segment boundary $\tau_j$ and summing the results:\[S = \sum_{j=1}^{N_{\mathrm{seg}}-1} \mathbb{I}(x \ge \tau_j).\]For instance, the example input $x=0.22$ is compared with the boundary $\tau_1 = 0.5$. Because $x < 0.5$, the indicator sum yields $S = 0$, indicating that the input falls into the $0$-th segment.

\textbf{Step 2: Local Interval Index Computation.} Given the segment index $S$, the local interval index $I$ within that specific segment is computed via a linear mapping, identical to the single-LUT approach:     
\begin{equation*}
    I = \left\lfloor \frac{x - \tau_{S}}{\tau_{S+1} - \tau_{S}} \cdot N_{\mathrm{br}} \right \rfloor = \left\lfloor \frac{0.22 - 0.1}{0.5 - 0.1} \cdot 4 \right \rfloor = \lfloor 1.2 \rfloor = 1
\end{equation*}
Thus, the example input $x=0.22$ maps to the local Interval 1 (representing the sub-range $[0.2, 0.3)$) within Segment 0.

\textbf{Step 3: Linear Evaluation.} Using the derived indices $(S, I)$, the corresponding linear coefficients $(a_{SI}, b_{SI})$ are fetched to compute the approximation $f(x) \approx a_{SI} \cdot x + b_{SI}$. In the running example, fetching the parameters for Segment $0$ and Interval $1$ yields $a_{01} = -16.327$ and $b_{01} = 8.163$. This produces $f(0.22) \approx -16.327 \cdot 0.22 + 8.163 = 4.571$, close to the true value $4.545$. The relative error drops to $0.57\%$, significantly outperforming the $34.8\%$ baseline.

\subsection{Cryptographic Lookup Protocol}
\label{sec:online-execution-protocol}

We now describe how to realize the three-step lookup under FHE. Step~1 is direct: each boundary comparison $x \ge \tau_j$ is evaluated via a homomorphic comparison primitive and the results are summed. However, Steps~2 and~3 pose the main challenge: both steps require multiplying the ciphertext $\llbracket x \rrbracket$ with real-valued parameters that depend on the segment index $S$ and local interval index $I$, such as the linear coefficients $(a_{SI}, b_{SI})$ in Step~3.  Since both indices are computed from the encrypted input $\llbracket x \rrbracket$, they also remain in ciphertext form as $\llbracket S \rrbracket_{\mathrm{LWE}}$ and $\llbracket I \rrbracket_{\mathrm{LWE}}$, and the server cannot directly extract the needed coefficients as in the cleartext setting. A na\"ive solution would use PBS to look up the required parameters with $\llbracket S \rrbracket_{\mathrm{LWE}}$ and $\llbracket I \rrbracket_{\mathrm{LWE}}$, but this yields them as LWE ciphertexts (e.g., $\llbracket a_{SI} \rrbracket_{\mathrm{LWE}}$). The subsequent multiplication $\llbracket x \rrbracket_{\mathrm{LWE}} \cdot \llbracket a_{SI} \rrbracket_{\mathrm{LWE}}$ would then demand high-precision fixed-point ciphertext--ciphertext arithmetic, which TFHE does not natively support.

\begin{figure}[t]
    \centering
    \includegraphics[width=\columnwidth]{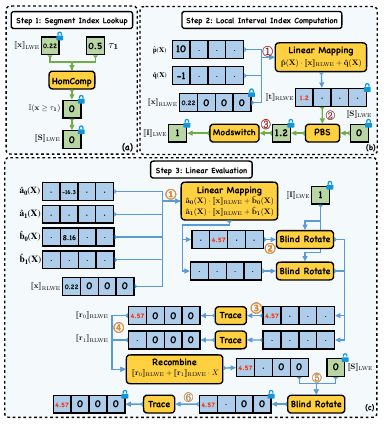}
    \Description{Three-stage dataflow for the cryptographic lookup example. It derives an encrypted segment index, computes a local interval index, and selects the corresponding linear slope and offset. Red entries trace the values selected for input 0.22.}
    \caption{\method{} cryptographic lookup on the running example ($N_{\mathrm{seg}}{=}2$, $N_{\mathrm{br}}{=}4$, $(\tau_0,\tau_1,\tau_2){=}(0.1,0.5,2.0)$, $x{=}0.22$). {\color{red}Red} marks the value selected at each step; ``$\cdot$'' marks entries omitted for clarity; the target LUT is \Cref{fig:segmented-lut}(d).}
    \label{fig:lookup-protocol}
\end{figure}

To overcome this, our protocol adopts an \emph{online-construct-then-select} strategy (Figure~\ref{fig:lookup-protocol}). Instead of first looking up the coefficients $(a_{SI}, b_{SI})$ and then computing $a_{SI} \cdot x + b_{SI}$, our protocol inverts the computation order: it first computes all candidate results via RLWE plaintext--ciphertext multiplication to construct a ciphertext test polynomial, then selects the correct one via PBS, thereby avoiding fixed-point arithmetic on LWE ciphertexts entirely. Concretely, the server encodes all coefficients $(a_{si}, b_{si})$ for every segment $s$ and interval $i$ into plaintext polynomials and multiplies them with the ciphertext $\llbracket x \rrbracket_{\mathrm{RLWE}}$, producing a ciphertext whose coefficients contain all candidate evaluations $a_{si} \cdot x + b_{si}$. The encrypted indices $\llbracket S \rrbracket_{\mathrm{LWE}}$ and $\llbracket I \rrbracket_{\mathrm{LWE}}$ are then used via PBS to select the correct result $a_{SI} \cdot x + b_{SI}$.

Based on this strategy, we now describe the protocol in detail, using the running example from \S\ref{sec:adaptive-segmented-lut} ($N_{\mathrm{seg}} = 2$, $N_{\mathrm{br}} = 4$, $(\tau_0, \tau_1, \tau_2) = (0.1, 0.5, 2.0)$, $x = 0.22$), as illustrated in Figure~\ref{fig:lookup-protocol}. The encrypted input is provided in two forms: an RLWE ciphertext $\llbracket x \rrbracket_{\mathrm{RLWE}}$ (with $x$ encoded in the constant coefficient) for the plaintext--ciphertext multiplications, and an LWE ciphertext $\llbracket x \rrbracket_{\mathrm{LWE}}$ for the homomorphic comparisons in Step~1. The preparation of both ciphertexts is detailed later.

\textbf{Step 1: Segment Index Lookup.} This step directly follows the cleartext algorithm, which evaluates $\llbracket S \rrbracket_{\mathrm{LWE}} = \llbracket \sum_{j=1}^{N_{\mathrm{seg}}-1} \mathbb{I}(x \ge \tau_j) \rrbracket_{\mathrm{LWE}}$. Each comparison $x \ge \tau_j$ is evaluated homomorphically using the high-precision comparison protocol \textsf{HomComp} from~\cite{bian2023he3db}, which internally chains several PBS operations for sufficient precision. The $N_{\mathrm{seg}} - 1$ comparison results, each an LWE ciphertext $\llbracket x \ge \tau_j \rrbracket_{\mathrm{LWE}}$, are then summed via homomorphic addition to produce the encrypted segment index $\llbracket S \rrbracket_{\mathrm{LWE}}$. This step requires $N_{\mathrm{seg}} - 1$ invocations of \textsf{HomComp} in total. In the running example (Figure~\ref{fig:lookup-protocol}(a)), a single $\textsf{HomComp}(x, \tau_1{=}0.5)$ is evaluated; since $x = 0.22 < 0.5$, the result is $\llbracket 0 \rrbracket_{\mathrm{LWE}}$, yielding $\llbracket S \rrbracket_{\mathrm{LWE}} = \llbracket 0 \rrbracket_{\mathrm{LWE}}$ (Segment~0).

\textbf{Step 2: Local Interval Index Computation.} As shown in Figure~\ref{fig:lookup-protocol}(b), we then homomorphically compute $\llbracket I \rrbracket_{\mathrm{LWE}}$ where $I = \lfloor \frac{x - \tau_S}{\tau_{S+1} - \tau_S} \cdot N_{\mathrm{br}} \rfloor$ via the online-construct-then-select strategy.

\emph{Construct.}
{\color{step2color}\textbf{\ding{192}}}~We precompute two plaintext polynomials, a slope $\hat{p}(X)$ and an offset $\hat{q}(X)$:
\begin{align*}
  \hat{p}(X) &= \textstyle\sum_{s=0}^{N_{\mathrm{seg}}-1} \frac{N_{\mathrm{br}}}{\tau_{s+1} - \tau_s} \cdot X^s
    &&\bigl(= 10.0 + 2.667\,X\bigr), \\
  \hat{q}(X) &= \textstyle\sum_{s=0}^{N_{\mathrm{seg}}-1} \Bigl({-}\frac{\tau_s \cdot N_{\mathrm{br}}}{\tau_{s+1} - \tau_s}\Bigr) \cdot X^s
    &&\bigl(= {-}1.0 - 1.333\,X\bigr).
\end{align*}
A single plaintext--ciphertext multiplication (\emph{linear mapping}) then constructs the \emph{test polynomial ciphertext}
\begin{equation*}
  \llbracket \mathbf{t} \rrbracket_{\mathrm{RLWE}} = \hat{p}(X) \cdot \llbracket x \rrbracket_{\mathrm{RLWE}} + \hat{q}(X).
\end{equation*}
Since $x$ is encoded in the constant coefficient of $\llbracket x \rrbracket_{\mathrm{RLWE}}$, multiplying by $\hat{p}(X)$ effectively broadcasts $x$ to every coefficient position. This produces a ciphertext whose $s$-th coefficient holds the candidate local interval index for segment~$s$:
\begin{equation*}
  \llbracket \mathbf{t} \rrbracket_{\mathrm{RLWE}}[s] = \frac{(x - \tau_s) \cdot N_{\mathrm{br}}}{\tau_{s+1} - \tau_s}
  \qquad \bigl(\text{coefficients: } [1.200,\; {-}0.747]\bigr).
\end{equation*}

\emph{Select and Round.}
{\color{step2color}\textbf{\ding{193}}}~A single PBS using $\llbracket S \rrbracket_{\mathrm{LWE}}$ then selects the $S$-th coefficient; in our example, coefficient~0 is selected, yielding the value $1.200$.
{\color{step2color}\textbf{\ding{194}}}~The \textsf{ModSwitch} operation then rounds this value, giving $\lfloor 1.200 \rfloor = 1$ and producing $\llbracket I \rrbracket_{\mathrm{LWE}} = \llbracket 1 \rrbracket_{\mathrm{LWE}}$ (Interval~1). In total, this step requires one plaintext--ciphertext multiplication, one plaintext addition, and one PBS.

\textbf{Step 3: Linear Evaluation.} As shown in Figure~\ref{fig:lookup-protocol}(c), the final step homomorphically computes $f(x) \approx a_{SI} \cdot x + b_{SI}$, where $S$ and $I$ are available only as LWE ciphertexts $\llbracket S \rrbracket_{\mathrm{LWE}}$ and $\llbracket I \rrbracket_{\mathrm{LWE}}$. Since the coefficients $(a_{SI}, b_{SI})$ depend on both encrypted indices, we apply the online-construct-then-select strategy twice, performing two rounds of selection, first by $I$, then by $S$:

\emph{Phase~1 Construct.}
{\color{step3color}\textbf{\ding{192}}}~For each segment $s$, we pack the slope and offset of every local interval $i \in \{0, \dots, N_{\mathrm{br}}-1\}$ from the per-segment LUTs (Figure~\ref{fig:segmented-lut}(f)) into a slope polynomial $\hat{a}_s(X) = \sum_{i} a_{si} \cdot X^i$ and an offset polynomial $\hat{b}_s(X) = \sum_{i} b_{si} \cdot X^i$. In the running example (Segment~0), $\hat{a}_0(X) = {-}47.06 - 16.33\,X - 8.25\,X^2 - 4.97\,X^3$ and $\hat{b}_0(X) = 14.12 + 8.16\,X + 5.77\,X^2 + 4.47\,X^3$. Using the same broadcast property as Step~2, a \emph{linear mapping} computes
\begin{equation*}
  \llbracket \mathbf{r}_s \rrbracket_{\mathrm{RLWE}} = \hat{a}_s(X) \cdot \llbracket x \rrbracket_{\mathrm{RLWE}} + \hat{b}_s(X),
\end{equation*}
so that the $i$-th coefficient of $\llbracket \mathbf{r}_s \rrbracket_{\mathrm{RLWE}}$ stores $a_{si} \cdot x + b_{si}$. In the running example, $\llbracket \mathbf{r}_0 \rrbracket$ has coefficients $(3.765,\; \mathbf{4.571},\; 3.959,\; 3.379)$ and $\llbracket \mathbf{r}_1 \rrbracket$ has $(2.538,\; 1.714,\; 1.296,\; 1.042)$, where the target $a_{01} \cdot 0.22 + b_{01} = 4.571$ sits at coefficient~1 of $\llbracket \mathbf{r}_0 \rrbracket$.

\emph{Phase~1 Select by $I$.}
Within each ciphertext, only the $I$-th coefficient holds the correct result.
{\color{step3color}\textbf{\ding{193}}}~A blind rotation using $\llbracket I \rrbracket_{\mathrm{LWE}}$ moves the $I$-th coefficient to the constant term.
{\color{step3color}\textbf{\ding{194}}}~A subsequent $\textsf{Trace}$ zeros out all other coefficients:
\begin{equation*}
  \llbracket \mathbf{r}_s \rrbracket \gets \textsf{Trace}\bigl(\textsf{BlindRot}(\llbracket \mathbf{r}_s \rrbracket_{\mathrm{RLWE}},\; \llbracket I \rrbracket_{\mathrm{LWE}})\bigr).
\end{equation*}
Each ciphertext now contains a single value $a_{sI} \cdot x + b_{sI}$ on the constant term, with the rest filled with zeros; in the running example, $\llbracket \mathbf{r}_0 \rrbracket = \llbracket 4.571 \rrbracket$ and $\llbracket \mathbf{r}_1 \rrbracket = \llbracket 1.714 \rrbracket$.

\emph{Phase~2 Construct (Recombine).}
{\color{step3color}\textbf{\ding{195}}}~To recombine the $N_{\mathrm{seg}}$ single-value ciphertexts into a single test polynomial ciphertext, we shift the $s$-th result into the $s$-th coefficient position via monomial multiplication and sum:
\begin{equation*}
  \llbracket \mathbf{R} \rrbracket_{\mathrm{RLWE}} = \textstyle\sum_{s=0}^{N_{\mathrm{seg}}-1} \llbracket \mathbf{r}_s \rrbracket \cdot X^s.
\end{equation*}
In the running example, the combined ciphertext stores $\mathbf{4.571}$ and $1.714$, with the target at position $S=0$.

\emph{Phase~2 Select by $S$.}
{\color{step3color}\textbf{\ding{196}}}~A final blind rotation using $\llbracket S \rrbracket_{\mathrm{LWE}}$ moves the $S$-th coefficient to the constant term.
{\color{step3color}\textbf{\ding{197}}}~A final $\textsf{Trace}$ isolates it:
\begin{equation*}
  \llbracket f(x) \rrbracket_{\mathrm{RLWE}} = \textsf{Trace}\bigl(\textsf{BlindRot}(\llbracket \mathbf{R} \rrbracket_{\mathrm{RLWE}},\; \llbracket S \rrbracket_{\mathrm{LWE}})\bigr) \approx \llbracket 4.571 \rrbracket_{\mathrm{RLWE}}.
\end{equation*}

In total, Step~3 requires $N_{\mathrm{seg}}$ plaintext--ciphertext multiplications, $N_{\mathrm{seg}}$ plaintext--ciphertext additions, $N_{\mathrm{seg}} + 1$ blind rotations, and $N_{\mathrm{seg}} + 1$ $\textsf{Trace}$ operations. Combined with Steps~1 and~2, the entire protocol uses $N_{\mathrm{seg}} - 1$ \textsf{HomComp} invocations, $N_{\mathrm{seg}} + 1$ plaintext--ciphertext multiplications, $N_{\mathrm{seg}} + 2$ blind rotations, and $N_{\mathrm{seg}} + 1$ \textsf{Trace} operations.

\subsection{Pipeline Integration}
\label{sec:pipeline-integration}

\begin{figure}[t]
    \centering
    \includegraphics[width=0.9\columnwidth]{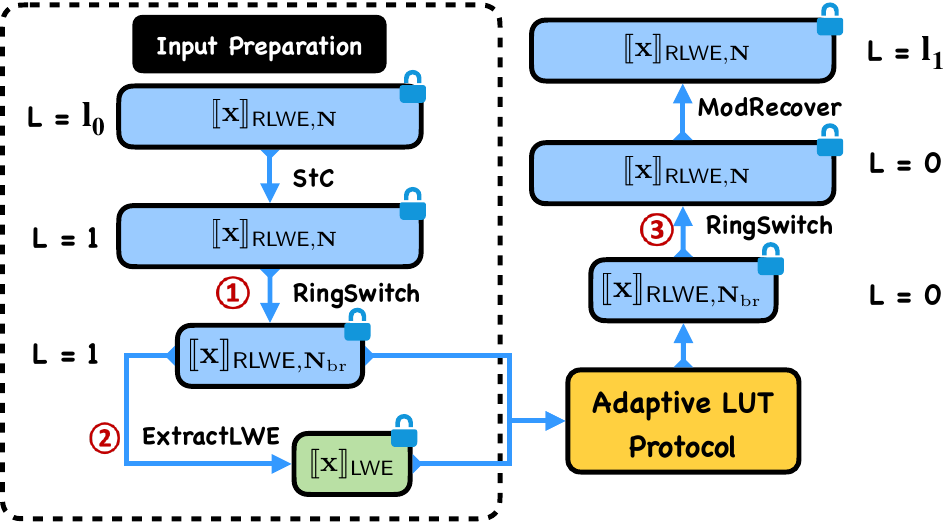}
    \Description{Pipeline converting a SIMD CKKS ciphertext to coefficient form, deriving small-ring RLWE and LWE inputs for the lookup protocol, and converting the lookup result back to a SIMD CKKS ciphertext.}
    \caption{Input/output pipeline: \textsf{StC} maps the CKKS ciphertext to coefficient form, then derive the RLWE and LWE ciphertexts with \textsf{RingSwitch} and \textsf{ExtractLWE}; after the lookup we return to CKKS via \textsf{RingSwitch} + \textsf{ModRecover}.}
    \label{fig:integration-protocol}
\end{figure}

With the lookup protocol in place, we now describe how to prepare its inputs and convert its output back to the SIMD domain (Figure~\ref{fig:integration-protocol}). We first describe the single-value pipeline (\S\ref{sec:pipeline-single}), and then extend it to the multi-value case (\S\ref{sec:pipeline-multi}).

\subsubsection{Input Preparation and Output Conversion.}
\label{sec:pipeline-single}

The lookup protocol requires the input $x$ in two ciphertext forms: an RLWE ciphertext $\llbracket x \rrbracket_{\mathrm{RLWE}}$ in coefficient form with $x$ encoded in the constant term, for the plaintext--ciphertext multiplications in Steps~2 and~3, and an LWE ciphertext $\llbracket x \rrbracket_{\mathrm{LWE}}$ for the homomorphic comparisons in Step~1. Since the preceding layers of the LLM pipeline produce SIMD-encoded ciphertexts, a scheme conversion is needed.

Starting from a SIMD-encoded ciphertext, we first apply \textsf{SlotToCoeff} (StC)~\cite{Cheon2018CKKS} to convert the slot representation ciphertext into coefficient form. The resulting coefficient-form ciphertext is then processed in two steps to produce the required inputs.
{\color{step2color}\textbf{\ding{192}}}~For the RLWE input, we apply \textsf{RingSwitch} to reduce the ring dimension from $N$ to $N_{\mathrm{br}}$, so that the subsequent polynomial multiplications in the lookup protocol operate on a much smaller ring at reduced cost.
{\color{step2color}\textbf{\ding{193}}}~We then apply \textsf{ModSwitch} to drop to level~0, then \textsf{ExtractLWE} to extract the constant term coefficient to get the LWE input. Both inputs are then sent into the lookup protocol.

After the lookup protocol completes, the output $\llbracket f(x) \rrbracket_{\mathrm{RLWE}}$ resides on the small ring of dimension $N_{\mathrm{br}}$ with the result in the constant term.
{\color{step2color}\textbf{\ding{194}}}~To return to the SIMD domain for subsequent layers, we first apply \textsf{RingSwitch} to embed the result back into the full ring of dimension $N$, then apply \textsf{ModRecover}, which internally performs \textsf{ModRaise}, \textsf{CoeffToSlot} (CtS), and \textsf{EvalMod}, to restore a valid CKKS ciphertext at the required level. Algorithm~\ref{alg:lookup-protocol} summarizes the complete protocol described above.

\subsubsection{Multi-Value Batching.}
\label{sec:pipeline-multi}

The protocol above targets the single-value case where one ciphertext carries a single element $x$. In certain scenarios, such as the reciprocal in Softmax normalization, $K$ elements (e.g., $K = 16$ attention heads) each require an independent non-linear evaluation. In this case, we perform \textsf{StC} only once to convert the SIMD ciphertext into coefficient form. After \textsf{StC}, the $K$ input values reside in $K$ coefficients of the coefficient-form ciphertext. We then apply \textsf{RingSwitch} to produce one or more small-ring ciphertexts of dimension $N_{\mathrm{br}}$; and each element lands at a known coefficient position in one of these ciphertexts.

To evaluate the $j$-th element ($0 \le j < K$), we take a copy of its corresponding small-ring ciphertext, multiply by the appropriate monomial $X^{-c}$ (where $c$ is the coefficient position of that element) to rotate it into the constant term, and apply a $\textsf{Trace}$ to zero out the remaining coefficients. The resulting ciphertext serves as the RLWE input $\llbracket x_j \rrbracket_{\mathrm{RLWE}}$ for the lookup protocol; the LWE input $\llbracket x_j \rrbracket_{\mathrm{LWE}}$ is obtained by applying \textsf{ModSwitch} and \textsf{ExtractLWE} to this same ciphertext. Both inputs are then fed into the lookup protocol. Once all $K$ evaluations complete, the output ciphertexts are recombined by shifting each result back to its original coefficient position and summing them. A single \textsf{RingSwitch} then maps the combined small-ring ciphertexts back into one full-ring ciphertext, and a single \textsf{ModRecover} restores it to the SIMD domain at the required level.

\begin{algorithm}[t]
\caption{\method{} Cryptographic Lookup Protocol}
\label{alg:lookup-protocol}
\begin{algorithmic}[1]
\REQUIRE CKKS ciphertext $\llbracket x \rrbracket_{\mathrm{CKKS}}$; precomputed polynomials $\hat{p}, \hat{q}, \{\hat{a}_s, \hat{b}_s\}_{s=0}^{N_{\mathrm{seg}}-1}$; boundaries $\{\tau_j\}_{j=0}^{N_{\mathrm{seg}}}$
\ENSURE CKKS ciphertext $\llbracket f(x) \rrbracket_{\mathrm{CKKS}}$
\STATE \textbf{// Input Preparation}
\STATE $\llbracket x \rrbracket_{\mathrm{coeff}} \gets \textsf{SlotToCoeff}(\llbracket x \rrbracket_{\mathrm{CKKS}})$
\STATE $\llbracket x \rrbracket_{\mathrm{RLWE}} \gets \textsf{RingSwitch}(\llbracket x \rrbracket_{\mathrm{coeff}},\; N_{\mathrm{br}})$
\STATE $\llbracket x \rrbracket_{\mathrm{LWE}} \gets \textsf{ExtractLWE}(\textsf{ModSwitch}(\llbracket x \rrbracket_{\mathrm{RLWE}}))$
\STATE \textbf{// Step 1: Segment Index}
\FOR{$j = 1$ \textbf{to} $N_{\mathrm{seg}} - 1$}
    \STATE $\llbracket x \ge \tau_j \rrbracket_{\mathrm{LWE}} \gets \textsf{HomComp}(\llbracket x \rrbracket_{\mathrm{LWE}},\; \tau_j)$
\ENDFOR
\STATE $\llbracket S \rrbracket_{\mathrm{LWE}} \gets \sum_{j=1}^{N_{\mathrm{seg}}-1} \llbracket x \ge \tau_j \rrbracket_{\mathrm{LWE}}$
\STATE \textbf{// Step 2: Local Interval Index}
\STATE $\llbracket \mathbf{t} \rrbracket_{\mathrm{RLWE}} \gets \llbracket x \rrbracket_{\mathrm{RLWE}} \cdot \hat{p}(X) + \hat{q}(X)$ \COMMENT{Construct}
\STATE $\llbracket I \rrbracket_{\mathrm{LWE}} \gets \textsf{PBS}(\llbracket \mathbf{t} \rrbracket_{\mathrm{RLWE}},\; \llbracket S \rrbracket_{\mathrm{LWE}})$ \COMMENT{select \& round}
\STATE \textbf{// Step 3: Linear Evaluation}
\FOR{$s = 0$ \textbf{to} $N_{\mathrm{seg}} - 1$ (construct)}
    \STATE $\llbracket \mathbf{r}_s \rrbracket_{\mathrm{RLWE}} \gets \llbracket x \rrbracket_{\mathrm{RLWE}} \cdot \hat{a}_s(X) + \hat{b}_s(X)$
\ENDFOR
\FOR{$s = 0$ \textbf{to} $N_{\mathrm{seg}} - 1$ (select by $I$)}
    \STATE $\llbracket \mathbf{r}_s \rrbracket \gets \textsf{Trace}(\textsf{BlindRot}(\llbracket \mathbf{r}_s \rrbracket,\; \llbracket I \rrbracket_{\mathrm{LWE}}))$
\ENDFOR
\STATE $\llbracket \mathbf{R} \rrbracket_{\mathrm{RLWE}} \gets \sum_{s=0}^{N_{\mathrm{seg}}-1} \llbracket \mathbf{r}_s \rrbracket \cdot X^s$ \COMMENT{Recombine}
\STATE $\llbracket f(x) \rrbracket_{\mathrm{RLWE}} \gets \textsf{Trace}(\textsf{BlindRot}(\llbracket \mathbf{R} \rrbracket,\; \llbracket S \rrbracket_{\mathrm{LWE}}))$ \COMMENT{Select by $S$}
\STATE \textbf{// Output Conversion}
\STATE $\llbracket f(x) \rrbracket_{\mathrm{CKKS}} \gets \textsf{ModRecover}(\textsf{RingSwitch}(\llbracket f(x) \rrbracket_{\mathrm{RLWE}}))$
\RETURN $\llbracket f(x) \rrbracket_{\mathrm{CKKS}}$
\end{algorithmic}
\end{algorithm}

%% file: docs/5-framework.tex
\section{Scheme-Aware Operator Selection}
\label{sec:framework}

As diverse nonlinear layers exist in Transformer models, determining which to evaluate with our TFHE-based segmented LUT protocol and which to keep in CKKS is non-trivial. To address this, we propose a scheme-aware operator selection method. We first categorize the operators inside the nonlinear layers and identify the subset that benefits from TFHE evaluation. We then propose an algorithm that jointly optimizes TFHE selection and level schedule.

\subsection{Operator Categorization}
\label{sec:framework-taxonomy}

As shown in Table~\ref{tab:taxonomy}, we classify the operators within the nonlinear layers of a Transformer model into two categories: \emph{arithmetic primitives} and \emph{non-arithmetic primitives}.

Arithmetic primitives include element-wise addition and multiplication, scalar addition and multiplication, and summation. For example, the LayerNorm affine $y = \gamma \cdot x + \beta$ is realized as a scalar multiplication followed by a scalar addition. These operations can be evaluated natively and efficiently by CKKS via its SIMD arithmetic: they are depth-cheap, fully parallelized across SIMD slots, and their cost is independent of the input value range. On the contrary, TFHE is not suitable for arithmetic primitives like multiplications. Therefore, all arithmetic primitives are assigned to CKKS by default, and the selection decision focuses on non-arithmetic primitives.

Non-arithmetic primitives include $\exp$, SiLU/GELU/ReLU, and $1/x$/$1/\sqrt{x}$. Under CKKS, these primitives are evaluated by a high-degree polynomial approximation or computed via an iterative method such as Goldschmidt iteration for $1/x$. Using CKKS for these primitives can be costly because such evaluations are \emph{level-hungry}: polynomial or iterative methods consume many multiplicative levels, so a single primitive may trigger frequent bootstraps. This makes non-arithmetic primitives the dominant source of cost in nonlinear layers, and makes them potential candidates for TFHE evaluation, since TFHE evaluates arbitrary functions directly through lightweight look-up tables and thus avoids the multiplicative-depth pressure of CKKS.

\begin{table}[t]
\centering
\footnotesize
\setlength{\tabcolsep}{4pt}
\caption{Definition of operators. Suffix \textsf{cc} denotes ciphertext--ciphertext operands; \textsf{cp} denotes ciphertext--plaintext.}
\label{tab:taxonomy}
\begin{tabular}{@{}l|l|l@{}}
\toprule
\textbf{Type} & \textbf{Name} & \textbf{Description} \\
\midrule
\multicolumn{3}{c}{\textbf{Arithmetic Primitives}} \\
\midrule
\multirow{2}{*}{Identity}
  & \textsf{ewadd\_cc/cp} & Element-wise addition of cc or cp \\
  & \textsf{ewmul\_cc/cp} & Element-wise multiplication of cc or cp \\
\midrule
\multirow{2}{*}{Expansion}
  & \textsf{sadd\_cp} & Scalar-broadcast addition \\
  & \textsf{smul\_cp} & Scalar-broadcast multiplication \\
\midrule
Reduction
  & \textsf{sum} & Summation along a specified dimension \\
\midrule
\multicolumn{3}{c}{\textbf{Non-Arithmetic Primitives}} \\
\midrule
\multicolumn{2}{l|}{\textsf{exp}}
  & Exponential $e^x$, used in Softmax \\
\midrule
\multicolumn{2}{l|}{\textsf{silu}, \textsf{gelu}, \textsf{relu}}
  & Non-linear activations in FFN layers \\
\midrule
\multicolumn{2}{l|}{\textsf{inv}, \textsf{invsqrt}}
  & Reciprocal $1/x$ and recip.\ sqrt $1/\sqrt{x}$ \\
\midrule
\multicolumn{3}{c}{\textbf{Composite Layers}} \\
\midrule
\multicolumn{2}{l|}{\textsf{Softmax}}
  & Attention score normalization with \textsf{exp} and \textsf{inv} \\
\midrule
\multicolumn{2}{l|}{\textsf{LayerNorm}}
  & Mean-variance normalization with \textsf{invsqrt} \\
\midrule
\multicolumn{2}{l|}{\textsf{RMSNorm}}
  & Root-mean-square normalization with \textsf{invsqrt} \\
\bottomrule
\end{tabular}
\end{table}

 Each nonlinear layer in a Transformer is a composition of arithmetic and non-arithmetic primitives. Some layers contain only a single non-arithmetic primitive (e.g., GELU and ReLU activation layers), while others are \emph{composite}, interleaving arithmetic primitives with non-arithmetic primitives. As illustrated in Figure~\ref{fig:framework}(a), the composite layer RMSNorm decomposes into a pipeline of seven primitives: given an input vector $x \in \mathbb{R}^d$, element-wise multiplication produces $x_i^2$, summation followed by scalar multiplication $\frac{1}{d}$ yields the mean of squares $\mu = \frac{1}{d}\sum_i x_i^2$, scalar addition gives $\mu + \epsilon$, reciprocal square root produces the normalization factor $(\mu + \epsilon)^{-1/2}$, and a final element-wise multiplication with scalar multiplication $\gamma$ yields the output $\gamma \cdot x \cdot (\mu + \epsilon)^{-1/2}$. In this pipeline, only the reciprocal square root $1/\sqrt{\cdot}$ is non-arithmetic; all remaining six primitives are arithmetic and evaluated natively by CKKS. 

This decomposition reveals that selecting an entire nonlinear layer for TFHE is unnecessarily costly, since the majority of its constituent primitives are arithmetic and already efficient under CKKS. Instead, we decompose each nonlinear layer into many primitives and restrict TFHE selection to non-arithmetic primitives alone, while all arithmetic primitives remain in CKKS. This fine-grained strategy reduces the scheme-assignment search space from all operators to only the few non-arithmetic candidates, and avoids wasting TFHE resources on operations that CKKS already handles well.

\subsection{Scheme-Aware Placement}
\label{sec:framework-placement}

The preceding analysis identifies non-arithmetic primitives as potential candidates for TFHE evaluation, but determining which ones to select is non-trivial because assigning a primitive to TFHE not only changes its own cost, but also affects the level assignment of the surrounding layers. First, selecting a non-arithmetic primitive for TFHE evaluation eliminates the multiplicative levels that its CKKS counterpart would have consumed, altering the overall level budget. Second, the input levels of the preceding and succeeding linear layers become additional decision variables that jointly determine both layer computation cost and scheme conversion cost; in particular, the higher the LWE-to-CKKS target level, the more expensive the conversion. Consequently, comparing the per-primitive cost of TFHE against CKKS in isolation is insufficient: the total cost depends on how the surrounding layers' levels are assigned, and the optimal level assignment in turn depends on how schemes are selected.

\begin{figure}[t]
\centering
\includegraphics[width=\columnwidth]{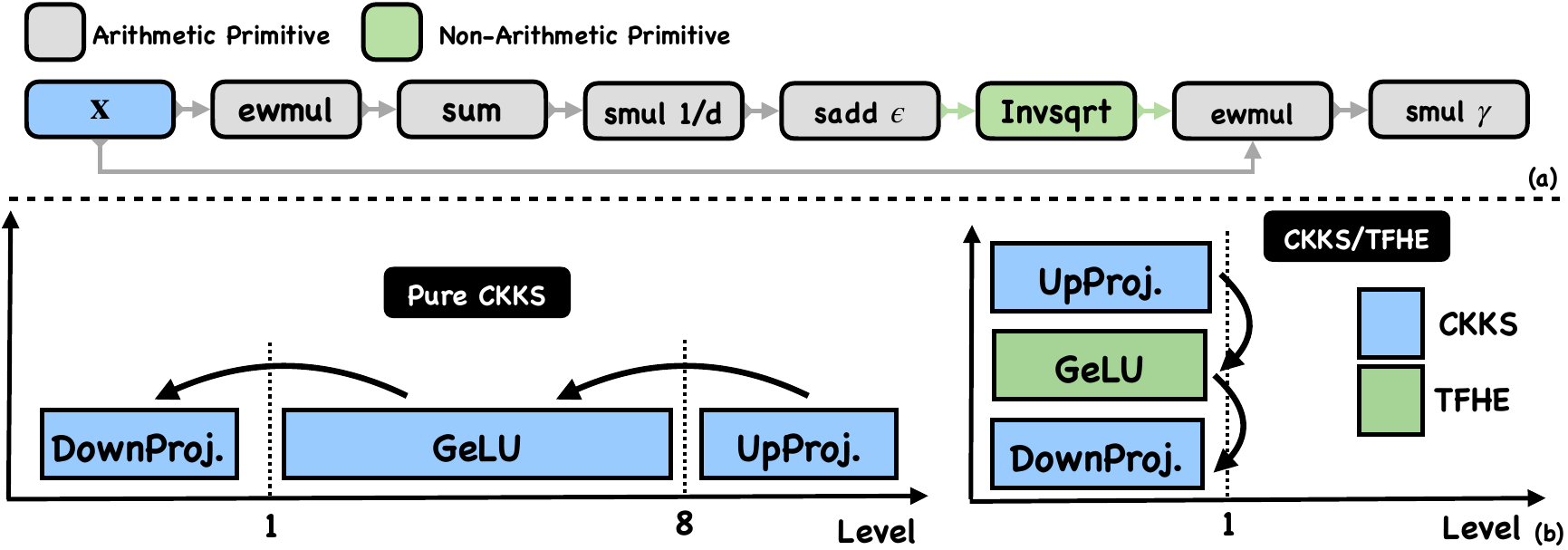}
\Description{Two-panel framework illustration. The first panel decomposes RMSNorm into arithmetic primitives and one reciprocal-square-root candidate for TFHE. The second contrasts CKKS level allocation for a GELU feed-forward network with the allocation obtained when GELU is evaluated using TFHE.}
\caption{(a) Primitive decomposition of RMSNorm, where only \textsf{invsqrt} (green) is a TFHE candidate. (b) Level management of a GELU FFN: pure CKKS (left) vs. TFHE selection (right).}
\label{fig:framework}
\end{figure}

We take GELU inside a Transformer FFN as an example (Figure~\ref{fig:framework}(b)). The FFN consists of an up-projection, a GELU activation, and a down-projection. Under pure CKKS, the up-projection is assigned a high input level to leave enough levels for the subsequent GELU evaluation. If GELU is instead selected for TFHE, the up-projection only needs enough levels for its own computation, since GELU no longer consumes any CKKS levels; meanwhile, the TFHE output only needs to be recovered to a level sufficient for the down-projection, which is exactly the LWE to CKKS target level that determines the conversion cost. Whether selecting GELU for TFHE thus requires jointly considering TFHE evaluation cost, conversion cost, and cost of other layers.

\begin{figure}[t]
\centering
\includegraphics[width=\columnwidth]{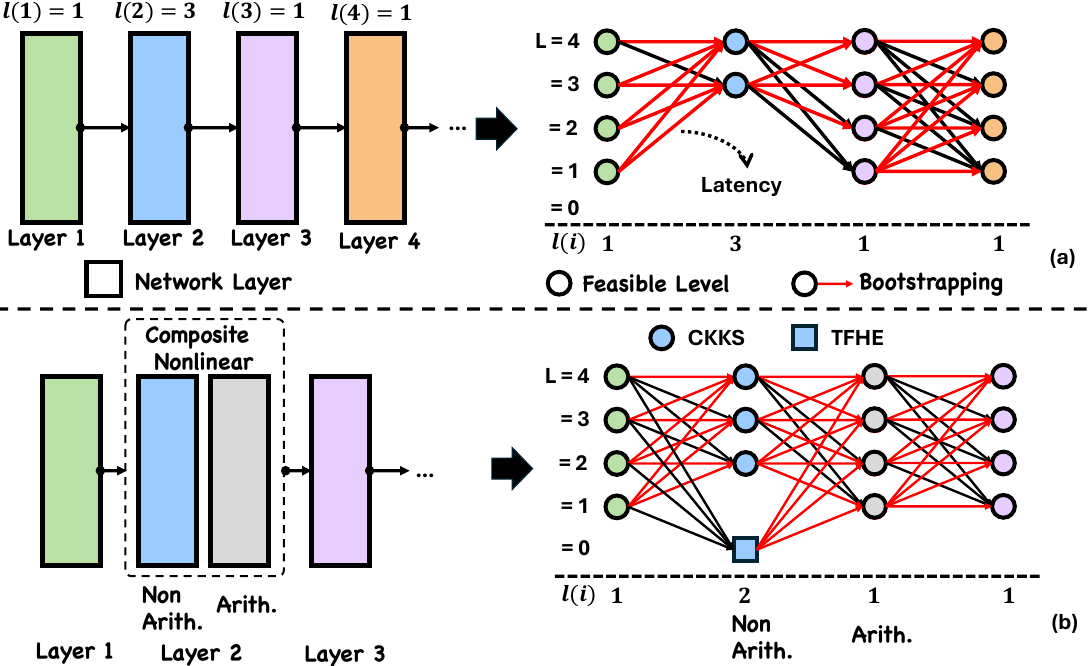}
\Description{Two directed acyclic graphs for operator placement. The pure-CKKS graph represents feasible ciphertext levels and a shortest level-allocation path. The scheme-aware graph adds a level-zero TFHE node for each nonlinear candidate so that the shortest path jointly selects the encryption scheme and CKKS levels.}
\caption{(a) Pure-CKKS level graph with the shortest path giving the optimal level schedule. (b) Scheme-aware extension: a TFHE node (blue, level~0) is injected at each non-arithmetic primitive, and the shortest path automatically chooses between the CKKS and TFHE paths.}
\label{fig:placement}
\end{figure}

To address this challenge, we build on CacheMir~\cite{yu2026cachemir}, a state-of-the-art level management algorithm designed for pure-CKKS private inference, and extend it to be scheme-aware. CacheMir formulates the problem as a shortest-path problem over a Directed Acyclic Graph (DAG) derived from the network structure. As shown in Figure~\ref{fig:placement}(a), CacheMir constructs a DAG by mapping each feasible input level of every layer to a vertex, and connecting the vertices of adjacent layers with edges whose weights capture the latency of computing the layer at the given input level plus any bootstrap needed to reach the next layer's input level. Since each layer is computed before any optional bootstrap or level drop~\cite{ebel2025orion}, input levels smaller than the layer's multiplicative depth are pruned as infeasible. Under this construction, a shortest path through the DAG corresponds to a globally optimal per-layer level assignment, which in turn gives the bootstrap placement. Formally, for a network with $D$ layers and a maximum level $L$, let $v[i,j]$ denote the vertex corresponding to the $i$-th layer with input level $j$, and let $e[i,x,y]$ denote the directed edge from $v[i,x]$ to $v[i+1,y]$ with weight $w[i,x,y]$. Let $\ell(i) \le L$ denote the multiplicative depth of layer $i$; an edge $e[i,x,y]$ is feasible only when $x \ge \ell(i)$. The minimum inference latency is then
\[
\min_{\{x_i\}} \;\; \sum_{i=1}^{D} w[i,\, x_i,\, x_{i+1}],
\]
where $\{x_i\}$ is the sequence of input levels chosen along the shortest path. Given that the latency of level drop is negligible and that bootstrap performance is insensitive to the input level~\cite{ebel2025orion}, the edge weight decomposes as
\[
w[i,\, x,\, y] \;\approx\; t_i(x) \;+\; \mathbf{1}_{x-\ell(i) < y} \cdot t_{\mathrm{boot}}(y),
\]
where $t_i(x)$ is the layer-$i$ computation latency at input level $x$ (monotonically increasing in $x$), $t_{\mathrm{boot}}(y)$ is the bootstrap latency to target level $y$ (also monotonically increasing), and the indicator equals $1$ when the residual level $x-\ell(i)$ after computation is insufficient for the desired output level $y$, triggering a bootstrap.

Building on this DAG formulation, our framework makes CacheMir scheme-aware in two steps.

\noindent\textbf{Step 1: Primitive-level decomposition.} Following \S\ref{sec:framework-taxonomy}, we expand each nonlinear layer in the DAG into its individual primitives. Each arithmetic primitive still becomes a standard CKKS vertex column in the level graph. Each non-arithmetic primitive is marked as a candidate for TFHE selection.

\noindent\textbf{Step 2: TFHE node injection.} For each tagged primitive $p$, we insert a special \emph{TFHE node} into the level graph alongside the existing CKKS vertices, as illustrated in Figure~\ref{fig:placement}(b). Since TFHE evaluation operates at level~0, this node is placed at level~0 in the graph; any vertex of the preceding layer can reach it via StC and ExtractLWE. Its outgoing edges connect to every feasible input level of the succeeding CKKS segment. A TFHE edge from $v[i,0]$ to $v[i+1,y]$ carries a weight of the form
\begin{align*}
w_{T}[i,\, 0,\, y] &\;=\; t_{\mathrm{StC}} \;+\; t_{\mathrm{TFHE}}(n_p,\, [\ell_p, r_p]) \;+\; t_{\mathrm{ModRecover}}(y) \\
&\;=\; t_{\mathrm{TFHE}}(n_p,\, [\ell_p, r_p]) \;+\; t_{\mathrm{boot}}(y),
\end{align*}
where the three terms correspond respectively to CKKS-to-LWE conversion, segmented-LUT evaluation, and LWE-to-CKKS conversion. The first term $t_{\mathrm{StC}}$ covers \textsf{SlotToCoeff} and \textsf{ExtractLWE}, and the second term $t_{\mathrm{TFHE}}(n_p, [\ell_p, r_p])$ is the TFHE evaluation cost governed by the segmented LUT configuration of \S\ref{sec:design}. Here, $n_p$ denotes the number of input elements processed by primitive $p$, and $[\ell_p,r_p]$ denotes its input range. The third term $t_{\mathrm{ModRecover}}(y)$ covers \textsf{ModRaise}, \textsf{CoeffToSlot}, and \textsf{EvalMod} to recover the output to level $y$. Since \textsf{SlotToCoeff} followed by \textsf{ModRecover} constitutes a full CKKS bootstrap (\textsf{SlotToCoeff} $+$ \textsf{ModRaise} $+$ \textsf{CoeffToSlot} $+$ \textsf{EvalMod}), we have $t_{\mathrm{StC}} + t_{\mathrm{ModRecover}}(y) = t_{\mathrm{boot}}(y)$, reusing a similar cost function from the pure-CKKS edge weight. The shortest path on this augmented DAG then gives the minimum-latency schedule, simultaneously determining which non-arithmetic primitives are evaluated by TFHE (those whose path traverses a TFHE node) and the optimal input level for every layer, without requiring a separate search over scheme selection decisions.

%% file: docs/5-evaluation.tex
\section{Evaluation}\label{sec:experiments}

\providecommand{\cell}[2]{\makecell{#1\\#2\,s}}

\begin{table}[!t]
\centering
\caption{Models used in evaluation.}
\label{tab:models}
\small
\begin{tabular}{@{}lrrr@{}}
\toprule
\textbf{Model} & \textbf{\#Heads} & \textbf{Hidden dim.} & \textbf{\#Layers} \\
\midrule
GPT-2 Base~\cite{radford2019gpt}            & 12 & 768  & 12 \\
TinyLlama-1.1B~\cite{tinyllama2024zhang}    & 16 & 2048 & 22 \\
LLaMA-3-8B~\cite{llama32024aaron} & 32 & 4096 & 32 \\
\bottomrule
\end{tabular}
\end{table}

\begin{table*}[!ht]
\centering
\caption{Softmax precision (bits, top) and latency (s, bottom) across context length $n$ and input range. ``--'' denotes unavailable or divergent results.}
\label{tab:softmax}
\small
\begin{tabular*}{\textwidth}{@{}c@{\hspace{8pt}}l|@{\extracolsep{\fill}}ccccc|ccccc|ccccc@{}}
\toprule
 & 
   & \multicolumn{5}{c|}{\textbf{Range $[-8,8]$}}
   & \multicolumn{5}{c|}{\textbf{Range $[-16,16]$}}
   & \multicolumn{5}{c}{\textbf{Range $[-32,32]$}} \\
\textbf{Cat.} & \textbf{Method}
                & 512 & 1k & 2k & 4k & 8k
                & 512 & 1k & 2k & 4k & 8k
                & 512 & 1k & 2k & 4k & 8k \\
\midrule
\multirow{3}{*}[-1ex]{\makecell[c]{Pure\\CKKS}}
& CacheMir
 & \cell{16.4}{149} & \cell{15.4}{149} & \cell{20.4}{149} & \cell{21.5}{149} & \cell{20.4}{150}
 & \cell{13.8}{149} & \cell{10.7}{149} & \cell{10.9}{149} & \cell{10.1}{149} & \cell{12.9}{149}
 & \cell{9.7}{234}  & \cell{8.4}{234}  & \cell{8.0}{234}  & \cell{9.5}{235}  & \cell{10.4}{236} \\
\cmidrule{2-17}
& NEXUS
 & \cell{14.7}{110} & \cell{15.6}{110} & \cell{16.6}{110} & \cell{17.6}{110} & \cell{18.6}{110}
 & -- & -- & -- & -- & --
 & -- & -- & -- & -- & -- \\
\cmidrule{2-17}
& MOAI
 & \cell{14.7}{110} & \cell{15.6}{110} & \cell{16.6}{111} & \cell{17.6}{113} & \cell{18.6}{117}
 & -- & -- & -- & -- & --
 & -- & -- & -- & -- & -- \\
\midrule
\multirow{2}{*}[-0.5ex]{\makecell[c]{Hybrid\\CKKS/TFHE}}
& PEGASUS
 & \cell{9.0}{69}  & \cell{9.0}{77}  & \cell{9.3}{94}  & \cell{10.5}{128} & \cell{9.2}{197}
 & \cell{6.7}{69}  & \cell{6.8}{77}  & \cell{7.9}{94}  & \cell{8.0}{128}  & \cell{8.6}{197}
 & \cell{5.9}{89}  & \cell{6.3}{97}  & \cell{7.0}{114} & \cell{7.6}{148}  & \cell{8.2}{217} \\
\cmidrule{2-17}
& \textbf{Ours}
 & \cell{\textbf{14.1}}{\textbf{32}} & \cell{\textbf{15.1}}{\textbf{31}} & \cell{\textbf{16.0}}{\textbf{31}} & \cell{\textbf{17.0}}{\textbf{31}} & \cell{\textbf{18.0}}{\textbf{32}}
 & \cell{\textbf{12.7}}{\textbf{51}} & \cell{\textbf{13.6}}{\textbf{51}} & \cell{\textbf{14.5}}{\textbf{51}} & \cell{\textbf{15.6}}{\textbf{51}} & \cell{\textbf{16.6}}{\textbf{51}}
 & \cell{\textbf{11.2}}{\textbf{70}} & \cell{\textbf{12.1}}{\textbf{70}} & \cell{\textbf{13.0}}{\textbf{70}} & \cell{\textbf{9.2}}{\textbf{71}}  & \cell{\textbf{15.1}}{\textbf{70}} \\
\bottomrule
\end{tabular*}
\end{table*}

\begin{table}[!t]
\centering
\caption{Wide-range Softmax comparison with Cho et al.~\cite{softmax2024cho} on $[-128,128]$. Each cell reports RMSE precision (bits, top) and latency (s, bottom).}
\label{tab:wide-softmax}
\small
\begin{tabular*}{\columnwidth}{@{}l@{\extracolsep{\fill}}ccccc@{}}
\toprule
\textbf{Method} & \textbf{512} & \textbf{1k} & \textbf{2k} & \textbf{4k} & \textbf{8k} \\
\midrule
Cho & \cell{17}{192} & \cell{16}{196} & \cell{9}{197} & \cell{9}{195} & \cell{4}{198} \\
\midrule
\textbf{Ours} & \cell{12}{117} & \cell{15}{118} & \cell{17}{120} & \cell{18}{121} & \cell{18}{122} \\
\bottomrule
\end{tabular*}

\end{table}

\begin{table*}[!t]
\centering
\caption{RMSNorm and LayerNorm precision (bits, top) and latency (s, bottom) across hidden dimension $d$ and variance $\sigma^2_{\max}$.}
\label{tab:norm}
\small
\begin{tabular*}{\textwidth}{@{}c@{\hspace{8pt}}l|@{\extracolsep{\fill}}ccc|ccc|ccc|ccc@{}}
\toprule
 & & \multicolumn{6}{c|}{\textbf{RMSNorm}}
   & \multicolumn{6}{c}{\textbf{LayerNorm}} \\
 & & \multicolumn{3}{c|}{$\sigma^2_{\max}\!=\!150$}
   & \multicolumn{3}{c|}{$\sigma^2_{\max}\!=\!8192$}
   & \multicolumn{3}{c|}{$\sigma^2_{\max}\!=\!150$}
   & \multicolumn{3}{c}{$\sigma^2_{\max}\!=\!8192$} \\
\textbf{Cat.} & \textbf{Method}
   & 768 & 2k & 4k & 768 & 2k & 4k
   & 768 & 2k & 4k & 768 & 2k & 4k \\
\midrule
\multirow{3}{*}[-1ex]{\makecell[c]{Pure\\CKKS}}
& CacheMir
 & \cell{10.4}{65} & \cell{10.6}{65} & \cell{11.0}{65}
 & \cell{11.2}{66} & \cell{10.4}{66} & \cell{9.1}{65}
 & \cell{8.8}{68}  & \cell{9.6}{68}  & \cell{10.7}{68}
 & \cell{6.7}{68}  & \cell{8.2}{68}  & \cell{10.7}{69} \\
\cmidrule{2-14}
& NEXUS
 & \cell{10.1}{148} & \cell{10.3}{148} & \cell{10.7}{148}
 & \cell{11.1}{189} & \cell{10.2}{189} & \cell{8.7}{189}
 & \cell{8.6}{151}  & \cell{9.4}{151}  & \cell{10.6}{151}
 & \cell{6.4}{191}  & \cell{7.9}{192}  & \cell{10.2}{192} \\
\cmidrule{2-14}
& MOAI
 & \cell{10.1}{148} & \cell{10.3}{149} & \cell{10.7}{151}
 & \cell{11.1}{189} & \cell{10.2}{190} & \cell{8.7}{192}
 & \cell{8.6}{151}  & \cell{9.4}{154}  & \cell{10.6}{157}
 & \cell{6.4}{192}  & \cell{7.9}{194}  & \cell{10.2}{198} \\
\midrule
\multirow{2}{*}[-0.5ex]{\makecell[c]{Hybrid\\CKKS/TFHE}}
& PEGASUS
 & \cell{6.9}{22} & \cell{6.4}{22} & \cell{6.6}{22}
 & \cell{4.6}{22} & \cell{4.7}{22} & \cell{4.7}{22}
 & \cell{8.7}{24} & \cell{8.9}{25} & \cell{8.6}{25}
 & \cell{6.9}{24} & \cell{6.9}{25} & \cell{6.9}{24} \\
\cmidrule{2-14}
& \textbf{Ours}
 & \cell{\textbf{14.1}}{\textbf{26}} & \cell{\textbf{11.9}}{\textbf{26}} & \cell{\textbf{14.1}}{\textbf{26}}
 & \cell{\textbf{11.2}}{\textbf{26}} & \cell{\textbf{14.2}}{\textbf{26}} & \cell{\textbf{15.8}}{\textbf{27}}
 & \cell{\textbf{14.6}}{\textbf{28}} & \cell{\textbf{14.6}}{\textbf{29}} & \cell{\textbf{14.6}}{\textbf{28}}
 & \cell{\textbf{12.5}}{\textbf{29}} & \cell{\textbf{10.1}}{\textbf{29}} & \cell{\textbf{14.9}}{\textbf{28}} \\
\bottomrule
\end{tabular*}
\end{table*}

\subsection{Experimental Setup}

\textbf{Implementation.}
\method{} is implemented with Lattigo~\cite{lattigo2024} as the CPU backend and a customized Phantom library~\cite{phantom2023yang} for GPU acceleration.
Since Lattigo does not expose a direct LWE API, we manage LWE ciphertexts with RLWE ciphertexts and construct our protocols on top of its blind-rotation primitive.
All baselines are re-implemented in the same framework with identical cryptographic parameters for fair comparison. We measure the latency of all methods at the granularity of modules to obtain a precise estimation of end-to-end runtime. All experiments are conducted on an Intel Xeon Gold 6240R CPU (2.40\,GHz, 48 threads) with 384\,GB of main memory, running Go\,1.24, CUDA\,12.4, and Lattigo v6. An NVIDIA A100 GPU (80\,GB) is also used for the end-to-end experiment in Table~\ref{tab:e2e}.

\textbf{Cryptographic Configuration.}
For the CKKS scheme, we set the polynomial degree to $N{=}2^{16}$ with a $1{,}763$-bit ciphertext modulus $Q$.
Following CacheMir~\cite{yu2026cachemir}, we set the maximum level to $L{=}13$ and the multiplicative depth of the bootstrapping circuit to $K{=}15$, adopting the Coefficients-to-Slots (CtS)-first bootstrapping variant with CtS depth $4$ and Slots-to-Coefficients (StC) depth $3$.
To support an 8-depth approximate modular reduction via a sparse secret key (Hamming weight $192$), we apply sparse secret encapsulation~\cite{sse2022jean}.
The RNS modulus chain is configured with $\log_2 q_0 \approx 53$ and $\log_2 q_i \approx 41$ for all $i{\ge}1$ to ensure sufficient noise budget.
For the TFHE scheme, we set the LWE dimension to $n_{\mathrm{LWE}}{=}1{,}024$ and the blind-rotation ring degree to $N_{\mathrm{br}}{=}2{,}048$, using a $53$-bit ciphertext modulus.
All parameter sets achieve 128-bit security according to the Homomorphic Encryption Standard~\cite{hestd2019albrecht}. For \method{}'s segmented LUT, we use $N_{\mathrm{seg}}{=}4$ segments per nonlinear primitive with linear interpolation for each segment. The boundaries are logarithmically spaced for $1/x$ and $1/\sqrt{x}$, and uniformly spaced for SiLU and GELU. We report precision in bits, defined as $-\log_2$ of the root-mean-square error between the FHE-evaluated output and the plaintext reference over inputs sampled from the target input range.

\textbf{Models, Datasets, and Baselines.}
We evaluate on three representative generative language models: GPT-2 Base~\cite{radford2019gpt}, TinyLlama-1.1B~\cite{tinyllama2024zhang}, and LLaMA-3-8B~\cite{llama32024aaron}, with key architectural details summarized in \Cref{tab:models}. Sequence lengths vary across experiments. We do not fine-tune any model and evaluate on several datasets. Perplexity is reported on several 8B models, which are most sensitive to errors introduced by FHE.

Our pure-CKKS baseline is CacheMir~\cite{yu2026cachemir}, which adopts the nonlinear operators from THOR~\cite{moon2025thor} without modification. Since THOR targets prefill, CacheMir is the directly comparable system for our decode-stage setting.
We additionally integrate the nonlinear methods of MOAI~\cite{zhang2025moai} and NEXUS~\cite{zhang2024nexus} as drop-in replacements for CacheMir's nonlinear operators. For hybrid CKKS/TFHE frameworks~\cite{lu2021pegasus,lohen2024} that share the same all-TFHE strategy, we use PEGASUS as a representative baseline.

\begin{table*}[!t]
\centering
\caption{End-to-end latency on CPU (hours, top) and GPU (minutes, bottom) across models and context lengths $n$.}
\label{tab:e2e}
\small
\providecommand{\ecell}[2]{\makecell{#1\,h\\#2\,m}}
\begin{tabular*}{\textwidth}{@{}c@{\hspace{8pt}}l|@{\extracolsep{\fill}}ccc|ccc|ccc@{}}
\toprule
 & & \multicolumn{3}{c|}{\textbf{GPT-2 Base}}
   & \multicolumn{3}{c|}{\textbf{TinyLlama-1.1B}}
   & \multicolumn{3}{c}{\textbf{LLaMA-3-8B}} \\
\textbf{Cat.} & \textbf{Method} & 512 & 2k & 8k & 512 & 2k & 8k & 512 & 2k & 8k \\
\midrule
Pure CKKS
& CacheMir
 & \ecell{1.58}{2.57} & \ecell{1.58}{2.57} & \ecell{1.60}{2.65}
 & \ecell{2.88}{4.97} & \ecell{2.88}{4.98} & \ecell{2.95}{5.45}
 & \ecell{4.25}{8.16} & \ecell{4.29}{8.39} & \ecell{4.44}{9.77} \\
\midrule
\multirow{2}{*}[-0.5ex]{\makecell[c]{Hybrid\\CKKS/TFHE}}
& PEGASUS
 & \ecell{1.08}{1.94} & \ecell{1.97}{3.83} & \ecell{5.70}{11.5}
 & \ecell{2.14}{4.20} & \ecell{4.33}{8.83} & \ecell{13.5}{27.8}
 & \ecell{4.24}{9.28} & \ecell{10.8}{23.0} & \ecell{37.5}{78.3} \\
\cmidrule{2-11}
& \textbf{Ours}
 & \ecell{\textbf{0.84}}{\textbf{1.31}} & \ecell{\textbf{0.84}}{\textbf{1.31}} & \ecell{\textbf{0.85}}{\textbf{1.39}}
 & \ecell{\textbf{1.52}}{\textbf{2.71}} & \ecell{\textbf{1.52}}{\textbf{2.71}} & \ecell{\textbf{1.57}}{\textbf{3.18}}
 & \ecell{\textbf{2.28}}{\textbf{5.18}} & \ecell{\textbf{2.31}}{\textbf{5.41}} & \ecell{\textbf{2.45}}{\textbf{6.79}} \\
\bottomrule
\end{tabular*}
\end{table*}

\begin{table*}[!t]
\centering
\caption{Relative PPL increase over the plaintext baseline (PPL$_0$) across models and datasets; lower is better.}
\label{tab:ppl}
\small
\setlength{\tabcolsep}{4pt}
\begin{tabular*}{\textwidth}{@{\extracolsep{\fill}}ll|rrrrrr@{}}
\toprule
\textbf{Model} & \textbf{Method}
  & \textbf{WikiText-2} & \textbf{WikiText-103} & \textbf{LAMBADA}
  & \textbf{GSM8K} & \textbf{ShareGPT} & \textbf{Avg.} \\
\midrule
\multirow{4}{*}{LLaMA-3-8B}
  & \textbf{Ours} & \textbf{+2.0\%} & \textbf{+2.2\%} & \textbf{+1.3\%}
  & \textbf{+0.7\%} & \textbf{+1.6\%} & \textbf{+1.6\%} \\
  & CacheMir & +2.0\% & +1.8\%
  & +1.3\% & +0.7\% & +1.1\%
  & +1.4\% \\
  & NEXUS/MOAI & -- & --
  & -- & -- & --
  & -- \\
  & PEGASUS & +41.4\% & +40.6\% & +23.0\% & +19.4\% & +25.8\% & +30.0\% \\
\midrule
\multirow{2}{*}{Qwen2-7B}
  & \textbf{Ours} & \textbf{+0.1\%} & \textbf{$-$0.0\%} & \textbf{+0.1\%}
  & \textbf{$-$0.1\%} & \textbf{+0.7\%} & \textbf{+0.1\%} \\
  & PEGASUS & +282.8\% & +278.7\% & +364.8\% & +196.8\% & +208.1\% & +266.2\% \\
\midrule
\multirow{2}{*}{Mistral-7B}
  & \textbf{Ours} & \textbf{+1.1\%} & \textbf{+0.8\%} & \textbf{+0.4\%}
  & \textbf{+0.4\%} & \textbf{+0.4\%} & \textbf{+0.6\%} \\
  & PEGASUS & +12.2\% & +12.5\% & +8.7\% & +8.3\% & +8.4\% & +10.0\% \\
\midrule
\multirow{4}{*}{DeepSeek-8B}
  & \textbf{Ours} & \textbf{+3.3\%} & \textbf{+3.3\%} & \textbf{+4.3\%}
  & \textbf{+2.4\%} & \textbf{+3.0\%} & \textbf{+3.3\%} \\
  & CacheMir & +0.4\% & +0.5\%
  & +0.8\% & $-$0.1\% & $-$0.2\%
  & +0.3\% \\
  & NEXUS/MOAI & -- & --
  & -- & -- & --
  & -- \\
  & PEGASUS & +166.2\% & +161.0\% & +139.1\% & +155.0\% & +167.5\% & +157.8\% \\
\midrule
\multirow{2}{*}{\textbf{All-model avg.}}
  & \textbf{Ours} & \textbf{+1.6\%} & \textbf{+1.6\%} & \textbf{+1.5\%}
  & \textbf{+0.9\%} & \textbf{+1.4\%} & \textbf{+1.4\%} \\
  & PEGASUS & +125.7\% & +123.2\% & +133.9\% & +94.9\% & +102.5\% & +116.0\% \\
\bottomrule
\end{tabular*}
\vspace{1pt}
\parbox{\textwidth}{\footnotesize -- indicates that PPL diverges.}
\end{table*}

\subsection{Micro-benchmarks} \label{sec:micro}

We first evaluate \method{} on three main nonlinear layers in LLMs, the Softmax, RMSNorm, and LayerNorm. After being processed by \method{}, each layer is decomposed into a combination of CKKS and TFHE operations based on its computational characteristics. We then evaluate several non-arithmetic primitives in isolation to further demonstrate the advantage of the segmented LUT design.

\textbf{Softmax.} \Cref{tab:softmax} shows the precision (bits) and latency (s) comparisons between \method{} and the baselines on Softmax. Following prior work~\cite{moon2025thor,yu2026cachemir}, inputs are generated from a normal distribution $\mathcal{N}(0,\sigma)$ truncated to range $[-M, M]$ with variance $\sigma^2 = M^2/3$ for $M \in \{8, 16, 32\}$, which approximates the distribution of attention scores in LLMs. For latency, \method{} is $2.2$--$4.8\times$ faster than the pure-CKKS baselines (CacheMir, NEXUS, MOAI), which all rely on iterative methods with many bootstraps, and $1.3$--$6.1\times$ faster than PEGASUS, whose latency grows with context length $n$ because it runs a per-element PBS for every exponential function. This shows that the hybrid design of \method{} can handle the heterogeneous workload of nonlinear operators and achieve better latency. For precision, \method{} is comparable to the pure-CKKS baselines on narrow ranges and gains a $2$--$5$-bit advantage on wide ranges where they degrade or fail to run, while consistently outperforming PEGASUS by $4$--$9$ bits, showing that the segmented LUT keeps precision robust as the input range widens.

To further evaluate Softmax over a wider input range, \Cref{tab:wide-softmax} compares \method{} with our reimplementation of Cho in Lattigo under identical cryptographic parameters on $[-128,128]$. Cho achieves $17$ and $16$ bits of RMSE precision at $n=512$ and $1\mathrm{k}$, but drops to $9$, $9$, and $4$ bits as $n$ increases to $2\mathrm{k}$, $4\mathrm{k}$, and $8\mathrm{k}$. In contrast, \method{} achieves $12$--$18$ bits across these context lengths and reaches $17$--$18$ bits at $n\geq2\mathrm{k}$, yielding an $8$--$14$-bit advantage in the long-context settings. It also reduces latency from Cho's $192$--$198$\,s to $117$--$122$\,s, corresponding to a $1.6\times$ speedup.

\textbf{RMSNorm and LayerNorm.} \Cref{tab:norm} compares \method{} with the baselines on RMSNorm and LayerNorm. Inputs are drawn from zero-mean normal distribution with per-group variance $\sigma^2$ sampled from two ranges $[0.2,\,150]$, and $[1.0,\,8192]$ following~\cite{moon2025thor}, across hidden dimensions $d\in\{768,\,2048,\,4096\}$. For latency, \method{} is $2.4$--$7.4\times$ faster than the pure-CKKS baselines (CacheMir, NEXUS, MOAI), which rely on iterative methods with frequent bootstraps, and achieves comparable latency to PEGASUS since both evaluate inverse square root on a single scalar with lightweight PBS. For precision, \method{} outperforms PEGASUS by $3$--$11$ bits across configurations, as the segmented LUT provides better precision over the full domain due to the linear approximation and adaptive method.

\textbf{Primitive Evaluation.}\label{sec:primitive}
We also benchmark non-arithmetic primitives $1/x$, $1/\sqrt{x}$, SiLU, and GELU to further demonstrate the performance of the segmented LUT design.

Table~\ref{tab:primitive-invx} reports results for $1/x$ and $1/\sqrt{x}$ over two different input ranges. For latency, \method{} is $8$--$9\times$ faster than Goldschmidt, which requires many iterations that consume multiple multiplicative levels and trigger bootstraps frequently. For precision, \method{} achieves $12.5$--$16.9$ bits and outperforms PEGASUS by $6$--$10$ bits, whose single-LUT approach is limited to $5$--$7$ bits.

\begin{table}[!t]
\centering
\caption{Single-primitive benchmark for $1/x$ and $1/\sqrt{x}$.}
\label{tab:primitive-invx}
\setlength{\tabcolsep}{4pt}
\small
\begin{tabular}{@{}llrrrrrr@{}}
\toprule
\textbf{Function} & \textbf{Range}
  & \multicolumn{2}{c}{\textbf{Goldschmidt}}
  & \multicolumn{2}{c}{\textbf{PEGASUS}}
  & \multicolumn{2}{c}{\textbf{Ours}} \\
\cmidrule(lr){3-4}\cmidrule(lr){5-6}\cmidrule(lr){7-8}
 & & Prec. & Lat.\,(s) & Prec. & Lat.\,(s) & Prec. & Lat.\,(s) \\
\midrule
$1/x$        & $[0.01, 10]$  &  9.8 & 45.5 &  5.4 & 0.8 & \textbf{12.7} & \textbf{4.8} \\
$1/x$        & $[0.01, 100]$ & 12.7 & 41.0 &  5.4 & 0.7 & \textbf{12.5} & \textbf{4.8} \\
$1/\sqrt{x}$ & $[0.01, 10]$  & 13.0 & 39.0 &  6.4 & 0.8 & \textbf{16.7} & \textbf{4.7} \\
$1/\sqrt{x}$ & $[0.01, 100]$ & 15.7 & 41.9 &  6.4 & 0.8 & \textbf{12.7} & \textbf{4.8} \\
\bottomrule
\end{tabular}
\end{table}

Results for SiLU and GELU on $[-20, 20]$ appear in Table~\ref{tab:primitive-activation}. \method{} is $2.7$--$2.9\times$ faster than a degree-59 Chebyshev approximation with comparable precision ($15.5$--$16.8$ bits vs.\ $12.4$--$15.2$ bits). PEGASUS achieves only $5.8$--$6.5$ bits, which \method{} outperforms by $9$--$10$ bits. Both benchmarks demonstrate the lightweight but accurate nature of the segmented LUT design, which can achieve high precision with a small number of segments and entries, while avoiding the heavy cost of iterative/polynomial methods.

\begin{table}[!t]
\centering
\caption{Activation function evaluation on $[-20, 20]$.}
\label{tab:primitive-activation}
\small
\begin{tabular*}{\columnwidth}{@{\extracolsep{\fill}}llrr@{}}
\toprule
\textbf{Function} & \textbf{Method} & \textbf{Prec.} & \textbf{Lat.\,(s)} \\
\midrule
SiLU & Chebyshev & 12.4 & 13.7 \\
SiLU & PEGASUS   &  5.8 &  0.8 \\
SiLU & \textbf{Ours} & \textbf{16.8} & \textbf{4.8} \\
\midrule
GELU & Chebyshev & 15.2 & 12.5 \\
GELU & PEGASUS   &  6.5 &  0.8 \\
GELU & \textbf{Ours} & \textbf{15.5} & \textbf{4.7} \\
\bottomrule
\end{tabular*}
\end{table}

We calibrate the evaluated input ranges by profiling the nonlinear inputs of LLaMA-3-8B with a fixed public prefix prepended~\cite{xiao2024streamingllm} across the five datasets used in our PPL evaluation. For Softmax, 31 of 32 layers have logit spans below 64 across all datasets, while the last layer reaches 105.5--109.2. Therefore, $[-32,32]$ captures the common case, and $[-128,128]$ covers the observed wide-range case. For RMSNorm, $\sigma^2_{\max}=8192$ covers approximately 72\% of the profiled normalization operators, representing the majority of practical cases. These measurements show that our benchmark ranges are representative of real-model inputs.

\subsection{End-to-End Performance}\label{sec:e2e}

To further demonstrate the effectiveness of \method{}, we present end-to-end latency across different models and context lengths. \Cref{tab:e2e} shows the end-to-end latency on CPU and GPU for GPT-2 Base, TinyLlama-1.1B, and LLaMA-3-8B at $n\in\{512, 2\mathrm{k}, 8\mathrm{k}\}$. Latency measures one-token generation (one decode iteration), excluding prefill. All methods share identical linear-layer implementations and differ only in nonlinear operators. Compared to CacheMir, \method{} achieves $1.5$--$2.1\times$ end-to-end speedup across all models and context lengths. NEXUS and MOAI are excluded as their Softmax implementation diverges at LLM-scale input ranges. Compared to PEGASUS, \method{} is $3$--$4\times$ faster, as PEGASUS relies on per-element PBS for Softmax whose cost scales linearly with $n$. On GPU, \method{} achieves $1.4$--$2.0\times$ speedup over CacheMir with a per-token latency of $1.3$--$6.8$\,m, demonstrating backend-agnostic gains and practical deployment potential.

\subsection{End-to-end Model Perplexity Evaluation}\label{sec:ppl}

We further evaluate end-to-end model quality with a ciphertext-aligned plaintext PPL test, where every nonlinear operation is replaced by its FHE counterpart. We test four 7--9\,B models (LLaMA-3-8B, Qwen2-7B, Mistral-7B, DeepSeek-R1-Distill-LLaMA-8B) on WikiText-2/103~\cite{wikitext2016merity}, LAMBADA~\cite{paperno2016lambada}, GSM8K~\cite{cobbe2021gsm8k}, and ShareGPT, comparing \method{} ($N_{\mathrm{seg}}{=}4$ segments) with PEGASUS. As reported in \Cref{tab:ppl}, \method{} keeps PPL degradation below $+4.3\%$ on every pair (average $+1.4\%$), while PEGASUS ranges from $+8\%$ to $+360\%$ (average $+116\%$). Such large degradations render outputs incoherent in practice: a single LUT cannot cover the wide input ranges of large LLMs, whereas our segmented LUT spans the full range with sufficient precision, validating the practicality of \method{}.
 We additionally evaluate CacheMir~\cite{yu2026cachemir}, NEXUS~\cite{zhang2024nexus}, and MOAI~\cite{zhang2025moai} on LLaMA-3-8B and DeepSeek-8B across all five datasets. CacheMir achieves PPL comparable to \method{} but incurs higher latency, whereas NEXUS/MOAI diverge because their fixed offline global maximum cannot cover LLM logit ranges.

%% file: docs/6-conclusion.tex
\section{Conclusion}
\label{sec:conclusion}

We propose \method{}, a hybrid CKKS/TFHE framework for efficient and accurate privacy-preserving LLM decoding, featuring an adaptive segmented LUT protocol for accurate nonlinear evaluation and a scheme-aware operator selector that jointly optimizes per-operator scheme assignment and CKKS-level allocation via shortest-path search. Experiments show that \method{} achieves significant speedup over prior frameworks with negligible PPL degradation, taking a meaningful step toward practical secure LLM inference.